\PassOptionsToPackage{style=ieee,backend=biber}{biblatex}

\documentclass[EPiC]{easychair}
\usepackage{amssymb,latexsym,amsfonts,amsmath,amsthm} 
\usepackage{biblatex} 
\usepackage{algorithmic} 
\usepackage{graphicx} 
\usepackage{xcolor} 
\usepackage{todonotes} 
\usepackage{multirow} 
\usepackage{makecell}
\usepackage{microtype} 
\usepackage{hyperref} 
\usepackage{cleveref} 
\usepackage[per-mode=symbol]{siunitx} 
\usepackage{xspace} 
\usepackage{csquotes} 
\usepackage{tabularx} 
\usepackage{booktabs}
\usepackage{siunitx}
\usepackage{colortbl} 
\usepackage{tikz}
\usepackage{subcaption}
\usepackage{carshapes}
\definecolor{forestgreen(web)}{rgb}{0.13, 0.55, 0.13}
\definecolor{deepskyblue}{rgb}{0.0, 0.75, 1.0}
\definecolor{flame}{rgb}{0.89, 0.35, 0.13}
\definecolor{brilliantrose}{rgb}{1.0, 0.33, 0.64}
\definecolor{chestnut}{rgb}{0, 0.4470, 0.7410}
\definecolor{darklavender}{rgb}{0.45, 0.31, 0.59}
\definecolor{darktangerine}{rgb}{1.0, 0.66, 0.07}
\definecolor{khakestari}{RGB}{205,205,205}
\definecolor{figureMainColor}{named}{blue} 
\tikzset{min_state/.style={draw, fill=figureMainColor,fill opacity=0.2,text opacity=1,  minimum height=5em, minimum width=4em, align=center,inner sep=5pt,rectangle, rounded corners}}
\tikzset{initial_state/.style={draw, fill=figureMainColor,fill opacity=0.2,text opacity=1,  minimum height=5em, minimum width=4em, align=center,inner sep=5pt,rectangle, rounded corners, anchor=north}}
\tikzset{initial_values/.style={draw=none, fill=none,  align=center, anchor=south}}

\usepackage[style=ieee,backend=biber]{biblatex}
\title{ARCH-COMP25 Category Report: Stochastic Models}
\author{
Alessandro Abate\inst{1} \and 
Omid Akbarzadeh\inst{2} \and
Henk A.P. Blom\inst{3} \and
Sofie Haesaert\inst{4} \and
Sina Hassani\inst{5} \and
Abolfazl Lavaei\inst{2} \and
Frederik Baymler Mathiesen\inst{3} \and
Rahul Misra\inst{5} \and
Amy Nejati\inst{2} \and
Mathis Niehage\inst{6} \and
Fie Ørum\inst{5} \and
Anne Remke\inst{6} \and
Behrad Samari\inst{2} \and
Ruohan Wang\inst{4} \and
Rafal Wisniewski\inst{5} \and
Ben Wooding\inst{2} \and
Mahdieh Zaker\inst{2}
}

\institute{
University of Oxford, Oxford, UK
\and
Newcastle University, Newcastle upon Tyne, UK
\and
Delft University of Technology, Delft, The Netherlands
\and
Eindhoven University of Technology, Eindhoven, The Netherlands
\and
Aalborg University, Aalborg, Denmark
\and
University of Münster, Münster, Germany
}
\authorrunning{Abate et. al.}
\titlerunning{ARCH-COMP25 Stochastic Models}

\newcommand{\tool}[1]{\ensuremath{\mathsf{#1}}}
\newcommand{\stochy}{\tool{StocHy}\xspace}
\newcommand{\syscore}{\tool{SySCoRe}\xspace}
\newcommand{\amytiss}{\tool{AMYTISS}\xspace}
\newcommand{\faust}{\tool{FAUST^2}\xspace}
\newcommand{\figaro}{\tool{FIGARO\ workbench}\xspace}
\newcommand{\srt}{\tool{SReachTools}\xspace}
\newcommand{\impact}{\tool{IMPaCT}\xspace}
\newcommand{\intervalmdp}{\tool{IntervalMDP.jl}\xspace}
\newcommand{\intervalmdpabstractions}{\tool{IntervalMDPAbstractions.jl}\xspace}
\newcommand{\prtct}{\tool{PRoTECT}\xspace}
\newcommand{\hypeg}{\tool{HYPEG}\xspace}
\newcommand{\hpnmg}{\tool{hpnmg}\xspace}
\newcommand{\prohver}{\tool{prohver}\xspace}

\newcommand{\pycatshoo}{\tool{PyCATSHOO}\xspace}

\newcommand{\modes}{\tool{modes}\xspace}
\newcommand{\modest}{\tool{Modest\ Toolset}\xspace}
\newcommand{\mascot}{\tool{Mascot\text{-}SDS}\xspace}
\newcommand{\hypro}{\tool{HyPro}\xspace}
\newcommand{\probreach}{\tool{ProbReach}\xspace}
\newcommand{\realyst}{\tool{RealySt}\xspace}

\newcommand{\sdcpn}{\tool{SDCPN\&IPS}\xspace}

\newcommand{\cpp}{\texttt{C++}\xspace}

\newcommand{\Y}{$\checkmark$}

\newcommand{\lightgrey}{black!10}
\newcommand{\rc}{\rowcolor{\lightgrey}}
\newcolumntype{Y}{>{\centering\arraybackslash}X}

\definecolor{grayfilling}{gray}{0.95} 

\begin{document}

\maketitle

\begin{abstract}
This report is concerned with a friendly competition for formal verification and policy synthesis of stochastic models. 
The main goal of the report is to introduce new benchmarks and their properties within this category and recommend next steps toward next year’s edition of the competition.
In particular, this report introduces \emph{three recently developed} software tools, a new water distribution network benchmark, and a collection of simplified benchmarks intended to facilitate further comparisons among tools that were previously not directly comparable.
This friendly competition took place as part of the workshop Applied Verification for Continuous and Hybrid Systems (ARCH) in Summer 2025.
\end{abstract}

\section{Introduction}\label{s:introduction}

The subgroup \enquote{Stochastic Models} of the annual friendly ARCH-Competition focuses on recent developments of tools that can analyze systems that exhibit uncertain, stochastic behavior.
This includes a diverse set of systems, expressing uncertainty in its various ways, \emph{e.g.,} continuously applied stochasticity or discrete mode changes, which happen with a certain probability.\vspace{0.2cm}

\noindent\fcolorbox{black}{white}{\parbox{.983\textwidth}{%
\textbf{Disclaimer} The presented report of the ARCH friendly competition for stochastic modeling group aims at providing a unified point of reference on the current state of the art
in the area of stochastic models together with the currently available tools and framework
for performing formal verification and optimal policy synthesis to such models. We further
provide a set of benchmarks, which we aim to use to push forward the development of current
and future tools. To establish further trustworthiness of the results, the code describing
the benchmarks together with the code used to compute the results is publicly available at \url{https://gitlab.com/goranf/ARCH-COMP}.
\smallskip
\\
This friendly competition is organized by
Abolfazl Lavaei (\small{\textsf{abolfazl.lavaei@newcastle.ac.uk}}),
Amy Nejati (\small{\textsf{amy.nejati@newcastle.ac.uk}}),
Anne Remke (\small{\textsf{anne.remke@uni-muenster.de}}),
and Alessandro Abate (\small{\textsf{alessandro.abate@cs.ox.ac.uk}}). 
}
}
\newpage
This report presents the results of the ARCH Friendly Competition 2025 in the group of \textit{stochastic models}. This group focuses on comparing tool developments for the analysis of complex systems that exhibit uncertain, stochastic behavior within Euclidean or hybrid state spaces~\cite{abate2024arch}. We refer the interested reader to the survey paper \cite{Lavaei_Survey} and references therein for the details of most of the underlying techniques used in the development of the tools of this category. To support these tool comparisons, the \textit{stochastic model} group develops relevant benchmarks and applies the participating tools to them. The inclusion of Euclidean state spaces distinguishes ARCH from the QComp, which compares tools analyzing stochastic processes evolving over discrete sets~\cite{Qcomp23}.

This report is organized as follows. Section~\ref{sec:toolsFrameworks} provides an overview of the tools and analysis frameworks participating in the stochastic models category. This includes newly introduced tools participating for the first time, listed in alphabetical order: \impact, \intervalmdp, and \prtct, as well as \emph{previously established} tools and analysis frameworks (also in alphabetical order): \amytiss, \faust, \figaro, \hpnmg, \hypeg, \mascot, the \modest, \probreach, \pycatshoo, \realyst, \sdcpn platform, \srt, \stochy{}, and \syscore. Section~\ref{sec:benchmarks} provides an overview of established benchmarks, \textit{i.e.,} stochastic models previously developed and applied by the stochastic models group. Section~\ref{sec:benchmarks1} discusses the development of new benchmarks, as well as extensions to existing ones. In particular, the collection of modeling and computational benchmarks has been enriched by a new \textit{water distribution network} benchmark. This benchmark can be used to check different specifications, primarily safety. Here safety is defined as whether the water level in the tank (state) is within prescribed bounds (safe set) for all time. As the system is subject to stochastic consumption disturbances, the goal is to verify the safety specification with probabilistic guarantee. Furthermore, we have introduced a collection of simplified examples aimed at enabling different tools to be applied with minimal modifications to the underlying model. The initiative for developing these benchmarks allows us to compare tools that previously were only applicable to distinct benchmarks. Section~\ref{sec:results} presents the 2025 results obtained for new tool-benchmark combinations, including relevant comparisons with prior benchmarking outcomes. Finally, Section~\ref{sec:conclusion} highlights key challenges and outlines future directions.

Similar to previous years, all participants were encouraged to provide a repeatability package (\textit{e.g.,} a Docker container) for centralized evaluation on the servers of the ARCH-group. Besides providing repeatable results, this allows for the sharing of the tools themselves to both the ARCH and the wider research community.

\section{Participating Tools \& Frameworks}\label{sec:toolsFrameworks}

The considered tools share the common setting in which the state space $\mathcal{X}$ is partitioned into three subsets: a target set $\mathcal{T}$, a safe set $\mathcal{S}$, and an unsafe set $\mathcal U = \mathcal{X} \setminus \mathcal{S}$. These tools and frameworks fall into two main categories: reachability assessment (a.k.a. verification) and control synthesis. 
The primary objective of reachability assessment tools is to compute/verify the probability of reaching specific subsets. A particularly challenging task is to estimate the probability of reaching unsafe sets that are \textit{rarely} reached, such as in the case of a mid-air collision between two passenger aircraft or a nuclear reactor meltdown.
The goal of control synthesis tools, however, is to synthesize a control law that ensures reaching the target set $\mathcal{T}$ while remaining within the safe set $\mathcal{S}$. Table~\ref{tab:main-objective} summarizes which tools and frameworks fall under which of these two categories. Most control synthesis tools are clearly capable of assessing reachability probabilities for non-rare events. In the sequel, we briefly introduce the main tools used in this report to obtain results.

\begin{table}[t!]
  \centering
  \caption{Main objective of participating tools/frameworks}
  \label{tab:main-objective}
  \begin{tabularx}{\linewidth}{@{}X X@{}}
    \toprule
    \textbf{Reachability Probability Assessment\footnotemark} & \textbf{Control Law Synthesis} \\
    \midrule
    \figaro & \amytiss\\
    \hpnmg & \faust\\
    \hypeg & \impact\\
    \modest & \intervalmdp\\
    \probreach & \mascot\\
    \prtct & \srt\\
    \pycatshoo & \stochy{}\\
    \realyst & \syscore\\
    \sdcpn & \\
    \bottomrule
  \end{tabularx}
\end{table}

\paragraph{PRoTECT.} The Python-based software tool \prtct~\cite{wooding2024protect,wooding2024protectposter} enables the parallelized construction of safety barrier certificates (BCs) for nonlinear polynomial systems. \prtct employs sum-of-squares (SOS) optimization programs~\cite{Yuan_SumOfSquares_py} to systematically search for polynomial-type BCs, aiming for the verification of safety properties across four classes of dynamical systems: \textit{(i)} discrete-time stochastic systems, \textit{(ii)} discrete-time deterministic systems, \textit{(iii)} continuous-time stochastic systems, and \textit{(iv)} continuous-time deterministic systems. Notably, \prtct is the first software tool that offers stochastic BCs, which correspond to categories~\textit{(i)} and~\textit{(iii)}~\cite{anand2020compositional,nejati2020compositional}, being relevant for ARCH. 
Implemented in Python, \prtct provides users with the flexibility of interaction either through its user-friendly graphical user interface (GUI) or programmatically via Python function calls as an application programming interface (API). At the same time, stochastic BCs can be synthesized either by specifying a minimum desired confidence level or by fixing a desired level set and optimizing the remaining variables to maximize the associated confidence. \prtct is compatible with the solvers \textsf{Mosek}~\cite{mosek} and \textsf{CVXOPT}~\cite{cvxopt}, and exploits parallelism across different BC degrees to efficiently identify a feasible one. The tool is publicly available at \url{https://github.com/Kiguli/PRoTECT}.

\footnotetext{Throughout the report, we also consider the verification problem within this category.}

\paragraph{IMPaCT.} Developed in \cpp,~\impact~\cite{wooding2024impactpaper,wooding2024impactposter} is designed for the parallelized verification and controller synthesis of large-scale stochastic systems using interval Markov chains (IMCs) and interval Markov decision processes (IMDPs), respectively. The tool serves to \textit{(i)} construct IMCs/IMDPs as finite abstractions of underlying original systems, and \textit{(ii)} leverage the interval iteration algorithm for formal verification and controller synthesis over (in)finite-horizon properties, including safety, reachability, and reach-while-avoid. It leverages \emph{interval iteration} algorithms to provide \emph{convergence guarantees} to an optimal controller in scenarios with infinite time horizons~\cite{haddad2018interval}. \impact can accommodate bounded disturbances and natively supports additive noises with different practical distributions, including normal and user-defined distributions provided by the custom PDF. \impact is designed using AdaptiveCpp~\cite{SYCL1,SYCL2}, an independent open-source SYCL implementation that enables adaptive parallelism across CPUs and GPUs from nearly all hardware vendors, including Intel and NVIDIA. \impact is flexible in the sense that any nonlinear optimization algorithm from the NLopt library can be used in the abstraction step. In addition, both abstractions and synthesized controllers can be imported from or exported to the standard \textsf{HDF5} data format~\cite{hdf5}. This enables synergy with a wide range of abstraction-based techniques, \emph{e.g.} from Gaussian processes~\cite{schon2024data}. The tool is publicly available at \url{https://github.com/Kiguli/IMPaCT}.

\paragraph{IntervalMDP.jl.} \intervalmdp, together with \intervalmdpabstractions, is a Julia-based toolbox for the verification and control synthesis of stochastic hybrid systems with respect to reachability, reach-while-avoid, safety, and expected exit time specifications. \intervalmdp is a package for defining Interval Markov Decision Processes (IMDPs), orthogonally decoupled IMDPs (odIMDPs), and mixtures of odIMDPS, and enables both verification and control synthesis via value iteration with a focus on flexibility, integration in pipelines, and optimal use of available hardware (CPU, GPU, memory) \cite{mathiesen2024intervalmdpjl, mathiesen2024scalable}. \intervalmdpabstractions builds on \intervalmdp to construct formal IMDP and odIMDP-based abstractions of discrete-time stochastic hybrid systems. The tool supports linear, smooth and piecewise nonlinear, and uncertain piecewise affine dynamics with additive (possibly non-diagonal) Gaussian or uniform noise, abstracted Gaussian processes \cite{jackson2021strategysynthesis}, and stochastic switched dynamics. \intervalmdpabstractions exploits structural properties of the system for tighter bounds on the satisfaction probability, less memory required, and faster computation of the abstraction and value iteration. Specification sets, \textit{e.g.,} reach or avoid regions, are described using \tool{LazySets.jl}, including hyper-rectangles, polytopes, zonotopes, and their unions, intersections, and Cartesian products. \intervalmdp is available at \url{https://github.com/Zinoex/IntervalMDP.jl} and \intervalmdpabstractions at \url{https://github.com/Zinoex/IntervalMDPAbstractions.jl}.

\paragraph{hpnmg.} 
The tool \hpnmg~\cite{huls2020hpmngtool} is a model checker for Hybrid Petri nets with an arbitrary but finite number of general transition firings against specifications formulated in signal temporal logic (STL)~\cite{huls2019ModelCheckingHPnGs}. 
Each general transition firing yields a random variable that follows a continuous probability distribution.
\hpnmg efficiently implements and combines algorithms for symbolic state-space construction~\cite{huls2019StateSpaceConstructionHybrid}, transformation into a geometric representation via convex polytopes~\cite{huls2018AnalyzingHybridPetri}, model checking of potentially nested STL formulas, and integration over the resulting satisfaction set to compute the probability that the specification holds at a specific time.
The tool is implemented in \cpp and utilizes the \hypro library~\cite{Schupp2017} for efficient geometric operations on convex polytopes, and the GNU Scientific Library (GSL) for multi-dimensional integration using Monte Carlo integration \cite{lepage1978NewAlgorithmAdaptive}. Different approaches to multi-dimensional integration have been compared with respect to scalability in \cite{huls2021StatespaceConstructionHybrid}. 
Recently, the tool has been extended with a guided simulation engine that applies statistical model checking (SMC) for a stochastic variant of STL to the precomputed symbolic state-space representation~\cite{Niehage23Valuetools,NiehageR25}.
Based on the algorithms developed for the tool \hypeg{}~\cite{pilch2017hypeg,pilch2017smc}, we also resolve discrete nondeterminism either probabilistically or via reinforcement learning, aiming to maximize or minimize the probability of a given property~\cite{niehage_learning_2021,Niehage2022Resilience}. This new method has also been extended to support rare-event simulation, offering a fully automated approach for stochastic hybrid models. An importance function is automatically derived that allows the use of importance splitting methods \emph{restart} and \emph{fixed effort} to efficiently identify rare events, which are expressed as STL properties~\cite{Niehage25NFM}.
The tool is available at \url{https://zivgitlab.uni-muenster.de/ag-sks/tools/hpnmg}.

\begin{table}[t!]
	\centering\small
	\caption{Tool-benchmark matrix: We indicate the year a
		tool was first applied to a given
		benchmark. Shortkeys: automated anesthesia (AS),
		building automation (BA), heated tank (HT), water
		sewage (WS), stochastic Van der Pol (VP),
		integrator chain (IC), autonomous vehicle (AV), patrol robot (PR), geometric Brownian motion (GB), minimal examples (ME), package delivery (PD), extended package delivery (PDx). The subscript $\underline{\cdot}$ and superscript $\overline{\cdot}$ indicate that a tool can handle a reduced version of a benchmark’s dynamics, but not the full benchmark, due to high-dimensional \emph{model} complexity and \emph{specification} complexity, respectively.
		\label{tab:resultsOverview}}
	\begin{tabularx}{\linewidth}{|l|Y|Y|Y|Y|Y|Y|Y|Y|Y|Y|Y|Y|} \hline
		\multirow{2}{*}{Tool} & \multicolumn{12}{c|}{Benchmarks} \tabularnewline\cline{2-13}
		& AS    & BA   & HT   & WS   & VP   & IC   &  AV    & PR & GB & ME & PD & PDx \tabularnewline\hline\hline
		\faust                & 2018  & 2018 &      &      &      & 2020 &    &     & & && \tabularnewline
		\rc{}\stochy{}        & 2019  & 2019 &      &      &      & 2020 &    &    & & && \tabularnewline
		\srt{}                & 2018  & 2018 &      &      &      & 2020 &    &     & & && \tabularnewline    
		\rc{}\amytiss         & 2020  & 2020 &      &      & 2020 & 2020 & 2020   & $\overline{2025}$ & 2021& & & \tabularnewline
		\hpnmg                &       &      &      & 2020 &      &      &    &    & & && \tabularnewline
		\rc{}\hypeg           &       &      & 2019 & 2020 &      &      &    &   &  & 2022 && \tabularnewline
		\mascot               &       &      &      &      & 2020 &      &    & 2021   & & && \tabularnewline          
		\rc{}\modes           &       &      & 2018 & 2020 &      &      &    &   & & 2022 && \tabularnewline
		\probreach            &       &      &      & 2020 &      &      &    &  & &  && \tabularnewline
		\rc{}\prohver         &       &      & 2020 & 2020 &      &      &    &   & & 2022 && 
		\tabularnewline
		\realyst         &       &      &      &      &      &      &    &   & & 2022 && 
		\tabularnewline
		\rc{}\sdcpn &       &      & 2019 &      &      &      &    &   &  2021 & & & \tabularnewline
		\syscore &   & 2021 &      &      &    2022  &      &   &   & & & 2022&2023 \tabularnewline
		\rc{}\figaro &       &      & 2021 &      &      &      &    &   & & && \tabularnewline
		\pycatshoo &   &  &  2021   &      &      &      &    & & & & & \tabularnewline 
		\rc{}\prtct &  \textcolor{black}{2025} & \textcolor{black}{2025} &     &      & \textcolor{black}{2025}     &  \textcolor{black}{2025}    &    & & & & & \tabularnewline
		\impact& \textcolor{black}{2025} & \textcolor{black}{2025} &     &      &      &    \textcolor{black}{$2025$}  &   \textcolor{black}{$\underline{2025}$} & \textcolor{black}{$\overline{2025}$} & & & \textcolor{black}{$2025$} & \tabularnewline
		\rc{}\intervalmdp & \textcolor{black}{2025} & \textcolor{black}{2025} &     &      &   \textcolor{black}{2025}  &   &   \textcolor{black}{$2025$} & \textcolor{black}{$\overline{2025}$} & & & & \tabularnewline\hline
	\end{tabularx}
\end{table}

\paragraph{SySCoRe.}
This toolset, short for Synthesis via Stochastic Coupling Relations, is a \texttt{MATLAB} toolbox designed to synthesize provably correct controllers for stochastic nonlinear (continuous-state) systems subject to possibly unbounded disturbances~\cite{huijgevoort2025syscore}. Given a system description and a co-safe temporal logic specification, \syscore provides all necessary functionality for synthesizing a robust controller and quantifying the associated formal robustness guarantees.
\syscore distinguishes itself from other tools by supporting both stochastic model order reduction techniques and state-space discretization, and by being applicable to nonlinear dynamics and complex co-safe temporal logic specifications over infinite horizons. To achieve this, \syscore constructs a finite abstraction—from a possibly reduced-order version of the system—and performs probabilistic model checking. It then establishes a probabilistic coupling between the original model and its finite abstraction, encoded as an approximate simulation relation, which is used to compute a lower bound on the specification satisfaction probability. 
A key feature of \syscore is that the computed error does not grow linearly with the specification horizon, enabling it to provide meaningful lower bounds even for infinite-horizon properties and unbounded disturbances.
Compared to the original version of \syscore~\cite{vanhuijgevoort2023syscore}, which was developed based on~\cite{haesaert2017verification}, the latest release~\cite{huijgevoort2025syscore} extends the toolset following the developments in~\cite{engelaar2023abstracting}, with a focus on \emph{stochastic model reduction}. In addition, several important strengths of the original version are retained and emphasized in this release, namely its comprehensiveness, computational efficiency, ease of use, and extensiveness.
\syscore~2.0 is publicly available at \url{https://github.com/BirgitVanHuijgevoort/SySCoRe-software}.  

\paragraph{AMYTISS.}
\texttt{AMYTISS}~\cite{AMYTISS,lavaei2020amytiss} is a software tool, implemented as a kernel on top of the acceleration ecosystem \texttt{pFaces}~\cite{khaled2019pfaces}, for designing correct-by-construction controllers of stochastic discrete-time systems. It implements parallel algorithms to (i) build finite Markov decision processes (MDPs) as finite abstractions of given original stochastic discrete-time systems~\cite{nejati2020compositional}, and (ii) synthesize controllers for the constructed finite MDPs satisfying bounded-time safety specifications and reach-avoid specifications~\cite{lavaei2024abstraction,lavaei2019automated}. The underlying computation parts are similar to the ones used in  \texttt{StocHy}~\cite{CauchiTACAS19}, and are used for compositional computations~\cite{lavaei2022dissipativity,lavaei2022scalable}. This tool significantly improves performances w.r.t. the computation time by parallel execution in different heterogeneous computing platforms including CPUs, GPUs and hardware accelerators (e.g., FPGA). In addition, \texttt{AMYTISS} proposes a technique to reduce the required memory for computing finite MDPs as on-the-fly abstractions (OFA). In the OFA technique, computing and storing the probability transition matrix are skipped. Instead, the required entries of the finite MDP are on-the-fly computed as they are needed for the synthesis part via the standard dynamic programming. This technique impressively reduces the required memory but at the cost of repeated computation of their entries in each time step from 1 to a finite-time horizon. This gives the user an additional control over the trade-off between the computation time and memory usage. The tool is publicly available at \url{https://github.com/mkhaled87/pFaces-AMYTISS}.

It is worth noticing that further tools participated in previous editions (see \textit{e.g.,}~\cite{ARCH23}).
\Cref{tab:resultsOverview} shows all tools that participated and the years in which they solved a benchmark the first time. We note that there was no ARCH competition for the stochastic modeling category in the year 2024.

\section{Established Benchmarks}\label{sec:benchmarks}
During previous ARCH editions, each of the two main tool/framework classes of Section~\ref{sec:toolsFrameworks} has stimulated the development of benchmarks that address the two corresponding challenges: reachability probability assessment and control law synthesis. These benchmarks include automated anesthesia (AS), building automation (BA), heated tank (HT), water sewage (WS), stochastic Van der Pol (VP), integrator chain (IC), autonomous vehicle (AV), patrol robot (PR), geometric Brownian motion (GB), minimal examples (ME), and package delivery (PD). Notably, some of these benchmarks cover different versions; this applies to HT, ME, and PD. Table~\ref{tab:benchmarks_new} shows the benchmarks that have been developed for each of the two aforementioned challenges, and for each benchmark, it also depicts the years during which results from competing tools have been produced and compared.
In Table~\ref{tab:resultsOverview}, we indicate the year a tool was first applied to a given benchmark. We also present an overview of benchmark properties in Table~\ref{tab:benchmarksOverview}. 

\begin{table}[t!]
		\centering
		\caption{Established benchmarks (in alphabetical sequence) for each challenge, and years when each benchmark has been applied in the comparison of competing tools.}
		\label{tab:benchmarks_new}
		\begin{tabularx}{\linewidth}{@{}>{\raggedright\arraybackslash}X
				>{\centering\arraybackslash}X
				>{\raggedright\arraybackslash}X
				>{\centering\arraybackslash}X@{}}
			\toprule
			\multicolumn{4}{c}{\textbf{Benchmark Challenge}} \\
			\midrule
			\multicolumn{2}{c}{\textbf{Reachability Probability Assessment}}
			& \multicolumn{2}{c}{\textbf{Control Law Synthesis}} \\
			\cmidrule(lr){1-2}\cmidrule(lr){3-4}
			\textbf{Benchmark} & \textbf{Results in}
			& \textbf{Benchmark} & \textbf{Results in} \\
			\midrule
			GB 
			& ’21 
			& AS 
			& ’18, ’19, ’20, ’21 \\
			HT 
			& ’18, ’19, ’20, ’21, ’22 
			& AV 
			& ’20, ’21 \\
			ME
			& ’22, ’23 
			& BA 
			& ’18, ’19, ’20, ’21, ’22 \\
			WS 
			& ’20, ’21 
			& IC
			& ’20, ’21 \\
			&      
			& PD
			& ’22, ’23 \\
			&      
			& PR 
			& ’21 \\
			&      
			& VP 
			& ’20, ’21, ’22 \\
			\bottomrule
		\end{tabularx}
	\end{table}

\begin{table}[htp]
     \centering\small
     \caption{Overview of benchmark properties. Shortkeys: Time horizon: Finite (F) or Infinite (I); Type of control: Switching (S), Drift (Dr), or Multiple (M); Time line: Discrete (D) or Continuous (C); State space: Continuous (C) or Hybrid (H); Drift in ODE/SDE: Linear (L), Piecewise Linear (pL), or Nonlinear (NL); Noise: State-dependent (SD) or fixed (FX), Gaussian (G) or Gaussian mixture model (GM), and Brownian motion (BM) or independently and identically distributed (iid), Rate/Size spont. jumps: State-dependent (SD) or fixed (FX)}
     \label{tab:benchmarksOverview}
     \begin{tabularx}{\linewidth}{|l|Y|Y|Y|Y|Y|Y|Y|Y|Y|Y|Y|Y|} \hline
                \multirow{2}{*}{Aspect} & \multicolumn{12}{c|}{Benchmarks} \tabularnewline\cline{2-13}
                                        & AS    & BA    & HT    & WS    & VP    & IC     & AV   & PR & GB & ME & PD & PDx \tabularnewline\hline\hline
          Liveness/deadlock             &       &       &       &       & \Y    &        &      & \Y  & & & &\tabularnewline
          \rc{}Prob. reachability       & \Y    & \Y    & \Y    & \Y    &       & \Y     & \Y   & & \Y &\Y & \Y &\Y \tabularnewline
          Control synthesis             & \Y    & \Y    &       &       &       & \Y     & \Y   & \Y && &\Y &\Y \tabularnewline
          \rc{}Min-max                  &       & \Y    &       &       &       &        & \Y   &  && & &\tabularnewline
          Time horizon                  &  F    &  F    &   F    &   F    & I     & F      & F    & I & F&F &I &I\tabularnewline
          \rc{}Type of control          &  S    &  M    &       &       & Dr    & Dr     & M    & M & & & &\tabularnewline
          Time line                     & D     & D     & C     & C     & D     & D      & D    & D & C&C &D &D\tabularnewline
          \rc{}State space              & C     & H     & H     & H     & C     & C      & C    & H & C&H &C &C\tabularnewline
          Drift in ODE/SDE              & pL    & NL    & NL    & pL    & NL    &  L     & NL   & NL& L&pL &L &L\tabularnewline
          \rc{}Noise in SDE             & FX & FX &       &       & FX & FX  & FX & FX & SD & SD & FX & FX\tabularnewline
          Noise: BM or i.i.d.           & iid   & iid   &       &       & iid   & iid    & iid  & iid & BM&  iid& iid  &iid\tabularnewline
           \rc{}Type of iid noise             &  &  &       &       &  &   &  & & &  & G &GM\tabularnewline
          Guards                   &       & \Y    & \Y    & \Y    &       &        & \Y   &  &&\Y & &\tabularnewline
          \rc{}Rate spont. jumps        & FX &       & SD & FX &       & FX  &      & & & SD & &\tabularnewline
          Size spont. jumps   & FX &       & FX & FX &       & FX  &      & & & FX & &\tabularnewline
          \rc{}Environment                   &       & \Y    &       & \Y    &       &        & \Y   & \Y && & &\tabularnewline
          Subsystems               &       & \Y    & \Y    & \Y    &       &        &      & & & & &\tabularnewline
          \rc{}Concurrency                   &       &       & \Y    & \Y    &       &        &      & && \Y& &\tabularnewline
          Synchronization          &       &       & \Y    & \Y    &       &        &      & && & &\tabularnewline
          \rc{}Shared variables              &       & \Y    &       & \Y    &       &        &      & && & &\tabularnewline
          \# discr. states       &       & 5     & 576   & 35    &       &        &      & 2 &&3-5 &1 &1\tabularnewline
          \rc{}\# continuous vars.       & 3     & 7     & 2     & 11    & 2     & 50     & 7    & 4 & 1&1-2 &2 &2\tabularnewline
          \# model params.      & 24    & 19    & 15    & 36    & 3     & 8      & 11   & 2 & 5& 7 &6 & 12\tabularnewline\hline
     \end{tabularx}
 \end{table}

\section{Extended Benchmarks}\label{sec:benchmarks1}
This year, we present new benchmarks, allowing for new outcomes by new tools. The details of each extended benchmark are presented below.

\subsection{Water distribution network} 
We introduce the first benchmark, \emph{i.e.,} the water distribution network, for the \emph{controller synthesis} task. It is worthwhile to note that the water distribution network consists of four main components: pumping stations, piping networks, water towers, and consumers. The dimension of the system depends on the number of elevated reservoirs and the number of pumping stations. For instance, we have considered a water distribution network with 2 pumping stations and 1 elevated reservoir; therefore, the dimension is 3. In the sequel, we describe the model in detail.
\subsubsection{Volume Balance} 
The water volume balance of the tower is modeled as
\begin{equation}
    \dot V(t)=\sum_{i=1}^{N_\text{q}} q_i(t) - \sum_{i=1}^{N_\text{d}}  d_i(t),
    \label{equ:mass_balance_water_tower}
\end{equation}
which captures the difference between the flow into and out of the tower. Moreover, in \eqref{equ:mass_balance_water_tower}, $V(t)$ denotes the volume of water in the tower at time $t$ [m$^3$], $N_\text{q}$ is the number of pump stations, and $q_i(t)$ represents the flow rate from pump station $i$ at time \( t \) [m$^3\,$s$^{-1}$]. Similarly, $N_\text{d}$ denotes the number of consumption groups, and $d_i(t)$ is the flow rate directed to consumption group $i$ at time $t$ [m$^3\,$s$^{-1}$].
The discrete-time model of \eqref{equ:mass_balance_water_tower}, obtained using the forward Euler discretization, is as
\begin{equation}
    V(t+t_\text{s})=V(t)+t_\text{s} \sum_{i=1}^{N_\text{q}} q_i(t)-t_\text{s}  \sum_{i=1}^{N_\text{d}} d_i(t),
    \label{equ:discrete-time-model}
\end{equation}
where, for the discretization to be exact, the consumption $d_i(t)$ and the pump station flow $q_i(t)$—both assumed to vary slowly—must remain constant over each sampling interval $t_\text{s}$.
It is of vital importance to prevent overflow in the water tower and to ensure that a minimum water reserve is maintained for emergency situations, such as firefighting.
These constraints on the minimum and maximum permissible water volume in the tower, denoted by \( \underline{V} \) and \( \overline{V} \), respectively, can be expressed as:
\begin{equation}
    \underline{V} \leq V(t) \leq \overline{V}.
    \label{equ:tower_constaint}
\end{equation}
\subsubsection{Pump Stations}\label{sec:pump_power_con}
The power consumption model of pump station $i$ is
\begin{equation}
    P_i(t)=\frac{1}{\eta_i}q_i(t)p_{i}(t),
    \label{equ:ideal_power_consumption_pump}
\end{equation}
where:
\begin{table}[h!]
    \begin{tabular}{l |l l }
        ~~~~~~~~~~~~~~~~~~~$P_{i}(t)$ & Power consumption of pump station $i$ at time $t$ & [W], \\
        ~~~~~~~~~~~~~~~~~~~$\eta_{i}$ & Efficiency of pump station $i$ & [$\cdot$], \\
        ~~~~~~~~~~~~~~~~~~~$p_{i}(t)$ & Pressure delivered by pump station $i$ at time $t$ & [Pa].
    \end{tabular}
\end{table}
\newline
The pressure delivered by the pump stations is determined as
\begin{subequations}
\label{equ:pipe_model}
\begin{align}
    p_{i} (t)&=\: \underbrace{r_{\text 
 f,i}|q_i(t)|q_i(t)}_\text{Pipe resistance} +\underbrace{r_{\text f,\Sigma}|q_\Sigma(t)|q_\Sigma(t)}_\text{Combined pipe resistance}+\underbrace{\rho_\text{w} g_0 h_\text{V}(t)}_\text{Water height} + \underbrace{\rho_\text{w} g_0 h_i}_\text{Tower elevation} \!\!\!\!\!\!\! ,\\
    q_\Sigma(t) &= \sum_{i=1}^{N_\text{q}}q_i(t) - \sum_{i=1}^{N_\text{d}}d_i(t), \\
    h_\text{V}(t) &= \frac{V(t)}{A_\text{t}},
    \label{equ:water_height}
\end{align}
\end{subequations}
where the individual pipe resistance, pressure from the elevation of the tower, and water height in the tower are considered. Of note is that the second term in \eqref{equ:pipe_model} becomes negative when the flow is directed out of the tower. In some studies, an additional term accounting for the kinetic energy of the water within the pipe is included. However, since the flow dynamics are stable and considerably faster than the dynamics of the tank, the flow quickly reaches a steady state. As a result, the kinetic energy term is neglected in \eqref{equ:pipe_model}. Moreover, all the variables used in \eqref{equ:pipe_model} are introduced in the following:
\begin{table}[h!]
    \begin{tabular}{l |l l }
        ~~~~~~~~~~~~~~~~~~~$r_{\text{f},i}$ & Pipe resistance  of pipe $i$  & [Pa s$^2$ m$^{-6}$], \\ 
        ~~~~~~~~~~~~~~~~~~~$\rho_\text{w}$ & Density of water  & [$997$ kg m$^{-3}$], \\ 
         ~~~~~~~~~~~~~~~~~~~$g_0$ & Gravitational acceleration & [$9.82$ m s$^{-2}$], \\
         ~~~~~~~~~~~~~~~~~~~$h_i$ & Pipe elevation & [m],\\
         ~~~~~~~~~~~~~~~~~~~$h_\text{V}(t)$ & Height of water inside the tower at time $t$ & [m],\\
         ~~~~~~~~~~~~~~~~~~~$A_\text{t}$ & Water surface area in cylindrical tower & [m$^2$]. 
    \end{tabular}
\end{table}

\!\!\!\!\!\!\!\!\!\! Combining \eqref{equ:ideal_power_consumption_pump} and \eqref{equ:pipe_model} results in the pump station power consumption
\begin{equation}
      P_i(t)=\frac 1 \eta_i q_i(t) \Big(r_{f,i}|q_i(t)|q_i(t)+r_{f,\Sigma}|q_\Sigma(t)|q_\Sigma(t)+\rho_w g_0\left(h_\text{V}(t) + h_i \right)\Big).
      \label{equ:pump_power}
\end{equation}
In practice, pump stations are subject to a total yearly extraction limit (TYEL), denoted by $\overline{Q}_i$, which represents the maximum volume of water that can be pumped over the course of a year to prevent overexploitation of the wells. In addition to this constraint, real-world pump stations have a maximum pump capacity $\overline q_i$ and do not have negative flow. This results in the following constraint: 
\begin{equation}
        0 \leq q_i(t) \leq \overline q_i.
        \label{equ:pump_constraint}
\end{equation}

\subsubsection{Constants and Constraints} \label{sec:Const_Const}
In the sequel, the constants and constraints to which the plant is subject are presented. Some values are based solely upon the physical constraints of the lab, while others are chosen. The utilized constants are shown in Table~\ref{tab:constant_model}. The elevations and pump station efficiency are chosen arbitrarily, given laboratory limitations. The cross-section of the water tower has been estimated based on its circumference.
\begin{table}[h!]
\caption{Constant values for the water distribution model.}
\label{tab:constant_model}
\centering
\vspace{0.1cm}
\begin{tabular}{@{}p{9.3cm}  c  c@{}}
    \toprule
    \textbf{Parameter} & \textbf{Symbol} & \textbf{Value} \\
    \midrule
    Cross‐section of water tower
      & $A_\mathrm{t}$ 
      & \SI{0.28}{\meter\squared} \\

    Elevation of tower compared to pump station  
      & $h_1$         
      & \SI{2.0}{\meter} \\

    Elevation of tower compared to pump station  
      & $h_2$         
      & \SI{1.5}{\meter} \\

    Pipe resistance
      & $r_{\mathrm{f},1}$  
      & \SI{0.35e5}{\pascal\,\hour\squared\meter^{-6}} \\

    Pipe resistance
      & $r_{\mathrm{f},2}$  
      & \SI{0.42e5}{\pascal\,\hour\squared\meter^{-6}} \\

    Pipe resistance        
      & $r_{\mathrm{f},\Sigma}$ 
      & \SI{0.29e5}{\pascal\,\hour\squared\meter^{-6}} \\

    Efficiency factor of pump station
      & $\eta_1$      
      & 0.90 \\

    Efficiency factor of pump station
      & $\eta_2$      
      & 0.80 \\
    \bottomrule
  \end{tabular}
\end{table}

The constraints for the plant are shown in Table~\ref{tab:constraints_system}. 
The TYEL has been chosen to be half the pump station's maximum flow. The water volume is chosen based on the volume of the tank. 
\begin{table}[h]
\caption{Constraints for the plant.}
\label{tab:constraints_system}
\centering
\begin{tabularx}{\linewidth}{@{}>{\raggedright\arraybackslash}X  c@{}}
    \toprule
    \textbf{Parameter} & \textbf{Value} \\
    \midrule
    Pump station flow 
      & $0 \le q_i(t) \le \SI{0.3}{\meter^3 \hour^{-1}}$ \\

    TYEL pump station 1 
      & $\SI{3.6}{\meter^3 \text{day}^{-1}}$ \\

    TYEL pump station 2  
      & $\SI{3.6}{\meter^3\text{day}^{-1}}$ \\

    Water volume 
      & $\SI{28}{\liter} \le V(t) \le \SI{155}{\liter}$ \\
    \bottomrule
  \end{tabularx}
\end{table}

\subsubsection{Water Consumption Model}
\label{sec:consumption_model}
\begin{figure}[h!]
    \centering
    \includegraphics[width=0.57\linewidth]{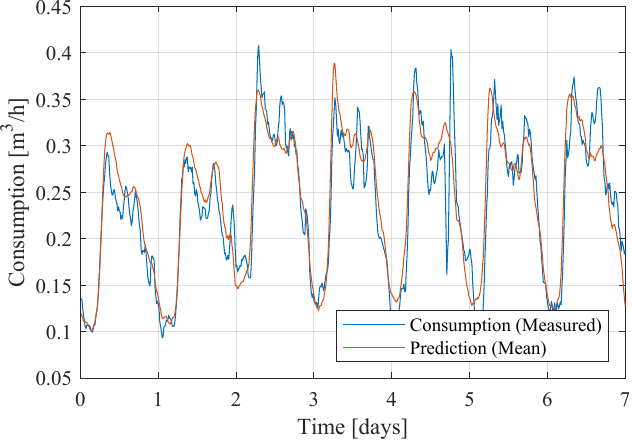}
    \caption{Current and predicted flow into the consumption group of the water distribution network in Bjerringbro, scaled to the operating range of the consumer valve controller.}
    \label{fig:consumption_model}
\end{figure}

Water consumption is the primary source of \emph{stochastic noise} in this model. A dataset of consumption from Bjerringbro Fælles Vandværk (Bjerringbro Water Distributor), supplying \num{3700} consumers, is available. The dataset consists of data for 88 days sampled every 15 minutes. Water consumption is known to be periodic within a day. However, differences between weekends and workdays are expected. For simplicity, the flow used in the consumption group is chosen to consist of the measured water consumption. The prediction model represents the mean flow into the consumption group for each 15-minute interval across the entire dataset. Both the prediction and consumption models have been scaled to match the operating range of the low-level controllers. The prediction and consumption profiles over a seven-day period are illustrated in Figure~\ref{fig:consumption_model}.
The figure reveals a notable correlation between the prediction and actual consumption, although discrepancies are evident, particularly at peak flow into the consumption group. 

\subsection{Lane Change Scenario}
\begin{figure}[t!]
		\centering
		\includegraphics[width=0.65\linewidth]{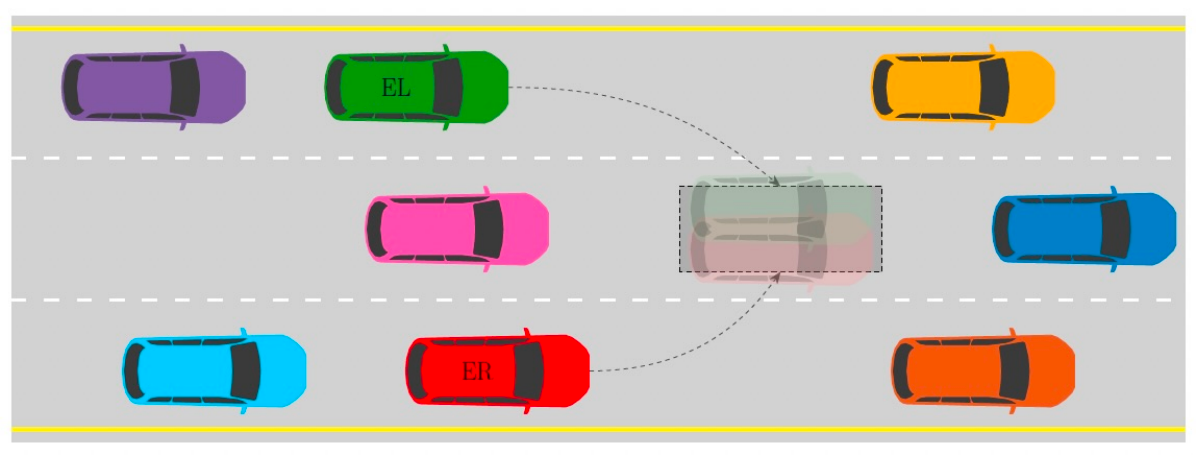}
		\caption{Lane-change scenario: The two AVs are denoted by $\mathcal{E} = \{EL, ER\}$, where the first letter signifies their role as ego vehicles and the second letter indicates their initial lane position—left or right, respectively.} 
		\label{fig:3lane}
	\end{figure}

Here, we present a lane-change scenario introduced and modeled as a general stochastic hybrid system (GSHS) in~\cite{zaker2024} to quantify the \emph{rare} collision probability of autonomous vehicles (AVs), which have situation awareness (SA). This scenario is depicted in Figure~\ref{fig:3lane} illustrating a lane-change maneuver involving two AVs, one green and one red with SA, initially positioned in the first and third lanes, respectively. The second lane contains an unoccupied gap amidst surrounding human-driven vehicles. At a particular time instant, both AVs independently decide to initiate a lane change. When the AV possessing SA detects the other AV’s intention to change lanes, it estimates the time-to-collision (TTC) and then uses it to assess the safety of the maneuver, guiding the AV to either continue with the lane change or return to its original lane.
    
We consider the following $5$D model, taken from \cite{pepy2006path}, for each ego vehicle $i\in\mathcal{E}$:
\begin{align}\notag
\mathsf{d}x_{t, i} & = (v_{x_{i}} \cos (\vartheta_{t, i}) - v_{y_{t, i}} \sin(\vartheta_{t,i}))\mathsf{d}t + \varepsilon_1 \mathsf{d}\mathbb P_t + \varepsilon_2 \mathsf{d}\mathbb{W}_t, \\\notag
\mathsf{d}y_{t, i} & = (v_{x_{i}} \sin (\vartheta_{t, i}) + v_{y_{t, i}} \cos(\vartheta_{t,i}))\mathsf{d}t + \varepsilon_1 \mathsf{d}\mathbb P_t + \varepsilon_2 \mathsf{d}\mathbb{W}_t, \\\notag
\mathsf{d}\vartheta_{t, i} & = \omega_{t, i}\mathsf{d}t, \\\notag
\mathsf{d}v_{y_{t, i}} \!\!& = (\frac{F_{yf}}{m}\cos(u_{t, i}) + \frac{F_{yr}}{m} - v_{x_{i}} \omega_{t, i}) \mathsf{d}t, \\\label{eq:vehicle model}
\mathsf{d}\omega_{t, i} & = (\frac{L_f}{I_z}F_{yf} \cos(u_{t, i}) - \frac{L_r}{I_z} F_{yr}) \mathsf{d}t.
\end{align}
In this model, $x_{t,i}$ and $y_{t,i}$ denote the coordinates of the vehicle's center of gravity in the longitudinal and lateral directions, respectively, while $\vartheta_{t,i}$ represents the vehicle’s orientation. The lateral velocity is given by $v_{y_{t,i}}$, whereas the longitudinal velocity $v_{x_i}$ is assumed to be constant. The yaw rate is denoted by $\omega_{t,i}$. Stochastic disturbances are modeled via a Poisson process $\mathbb{P}_t$ with intensity $\lambda_1$ and reset magnitude $\varepsilon_1$, and a Brownian motion $\mathbb{W}_t$ with diffusion coefficient $\varepsilon_2$. The sole control input is the front wheel steering angle $u_{t,i}$. To execute the lane-change maneuver, a PD controller given by
$
u_{t,i} = K_p \big(y_{d,i} - y_{t,i}\big) - K_d \frac{\mathrm{d}y_{t,i}}{\mathrm{d}t},
$
can be employed, where $y_{d,i}$ denotes the desired lateral position, and $K_p = 1.5 \times 10^{-3}$, $K_d = 10^{-2}$ are the proportional and derivative gains, respectively.
Under a linear tire model, the lateral forces on the front and rear tires, denoted by $F_{yf}$ and $F_{yr}$, are expressed as
$
F_{yf} = -C_{\alpha f} \alpha_f, \quad F_{yr} = -C_{\alpha r} \alpha_r,
$
where $C_{\alpha f}$ and $C_{\alpha r}$ are the cornering stiffness coefficients for the front and rear tires, respectively. The corresponding slip angles $\alpha_f$ and $\alpha_r$ are given by
$
\alpha_f = \frac{v_{y_{t,i}} + L_f \omega_{t,i}}{v_{x_i}} - u_{t,i}, \quad
\alpha_r = \frac{v_{y_{t,i}} - L_r \omega_{t,i}}{v_{x_i}},
$
with $L_f$ and $L_r$ denoting the distances from the center of gravity to the front and rear axles. The parameters in this model are specified as $v_{x_i} = 20\, \text{m/s}$, $\varepsilon_1 = 10^{-6}$, $\varepsilon_2 = 10^{-2}$, $\lambda_1 = 0.5$, $m = 2000\, \text{kg}$, $I_z = 2000\, \text{kg}\cdot\text{m}^2$, $C_{\alpha f} = C_{\alpha r} = 6 \times 10^4\, \text{N/rad}$, and $L_f = L_r = 2\, \text{m}$.

Following the GSHS definition \cite{bujorianu2006toward}, the continuous state vector of AV $i$ is defined as $\mathbf{x}_{t,i} =[x_{t,i}, y_{t,i}, \vartheta_{t,i}, v_{y_{t,i}}, \omega_{t,i})]$. The discrete state of the AV $ER$ is denoted by $\theta_{t, ER} \in \Theta_{ER} = \{(0, Off),\, (1, 1^-),\, (2, 1^-),\, (-1, 1^+),\, (Hit, \star)\},
$
and for AV $EL$, the discrete state is given by $\theta_{t, EL} \in \Theta_{EL} = \{(0, Off),\, (1, 1^+),\, (Hit, \star)\}.
$
In both sets, the symbol $\star$ represents a non-contributory component whose value does not influence the outcome. Each discrete state component conveys specific behavior or intent, interpreted as follows:
\begin{itemize}
	\item $0$: the AV is proceeding straight without initiating a lane change;
	\item $1$: the AV is actively changing lanes;
	\item $2$: the AV has detected that the other vehicle is attempting a lane change;
	\item $-1$: the AV is reversing its lane-change decision (i.e., returning to the original lane);
	\item $Hit$: a collision has occurred between the two AVs;
	\item $Off$: the AV's indicators are off, indicating no lane change is anticipated;
	\item $1^+$: the right indicator is flashing, and the AV is executing a lane change to the right;
	\item $1^-$: the left indicator is flashing, and the AV is executing a lane change to the left.
\end{itemize}
As can be noted, since $EL$ is assumed to lack SA, the modes $2$ and $-1$ are not applicable to its discrete state.

\begin{figure}[t!]
	\centering
	\includegraphics[width=0.35\linewidth]{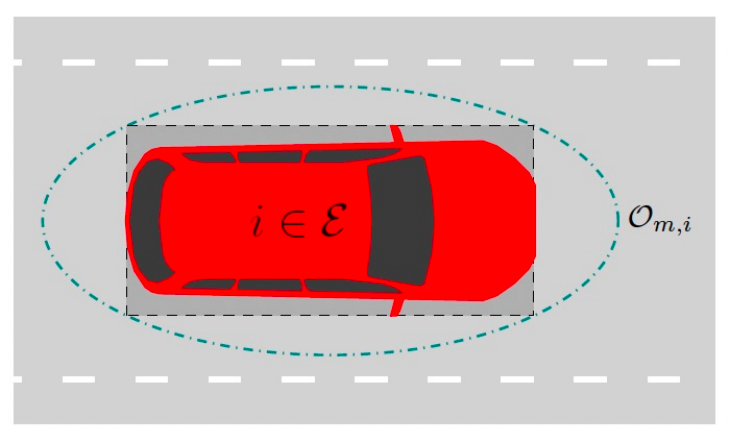}
	\caption{The circumscribed ellipse $\mathcal O_{m, i}$ around each AV.}
\label{fig:ellipse level sets}
\end{figure}

To assess potential collisions, we consider a circumscribed ellipse around each AV, as depicted in Figure~\ref{fig:ellipse level sets}. The ellipse associated with AV $i$ is defined by
$
\mathcal{O}_{m,i}: \frac{(x - x_{t,i})^2}{\mathcal{R}_x^2} + \frac{(y - y_{t,i})^2}{\mathcal{R}_y^2} = 1,
$
where $\mathcal{R}_x = \frac{\sqrt{2}}{2}l_v$ and $\mathcal{R}_y = \frac{\sqrt{2}}{2}w_v$ represent the semi-major and semi-minor axes, respectively, where $l_v = 4.508\, \text{m}$ and $w_v = 1.61\, \text{m}$ denote the vehicle's length and width, and $(x_{t,i}, y_{t,i})$, for $i \in \mathcal{E}$, specifies the center of the ellipse. A collision is declared to occur if the ellipses intersect, \emph{i.e.}, if $\mathcal{O}_{m,i} \cap \mathcal{O}_{m,j} \neq \emptyset$ for $i, j \in \mathcal{E}$, with $i \neq j$.

The AV $ER$ cannot receive information of AV $EL$ unless the two vehicles are sufficiently close for information exchange. To model this condition, we define another elliptical region 
$
\mathcal{O}_{SA,i}: \frac{(x - x_{t,i})^2}{\mu_{r_x}^2} + \frac{(y - y_{t,i})^2}{\mu_{r_y}^2} = 1,
$
around each AV as its awareness zone, where $\mu_{r_x}$ and $\mu_{r_y}$ denote the semi-axes of the awareness region centered at $(x_{t,i}, y_{t,i})$. When the awareness regions of two AVs intersect, \emph{i.e.}, $\mathcal{O}_{SA,i} \cap \mathcal{O}_{SA,j} \neq \emptyset$, AV $ER$ becomes aware of $EL$ and is able to receive the necessary state information.

Upon detecting that $\theta_{t, EL} = (1, 1^+)$, indicating that $EL$ is executing a right lane change, $ER$ requires a finite response time before updating its discrete state to $\theta_{t, ER} = (2, 1^-)$, which reflects its awareness of $EL$'s maneuver. This delay is captured by an instantaneous transition rate defined as
$
\lambda_2(\theta_{t, ER}, \mathbf{x}_{t, ER}) = \chi\big(\theta_{t, ER} = (1, 1^-)\big)\frac{p_{\text{delay}}(\eta)}{\int_{\eta}^{\infty} p_{\text{delay}}(s) \, \mathrm{d}s},
$
where $\chi(\cdot)$ is the indicator function, and $p_{\text{delay}}(s)= \frac{s}{\mu_d^2} e^{-s^2 / (2\mu_d^2)}$ is a Rayleigh probability density function modeling the reaction time with $\mu_d$ denoting the mean reaction delay.

\begin{figure}[t!]
	\centering
	\includegraphics[width=0.9\linewidth]{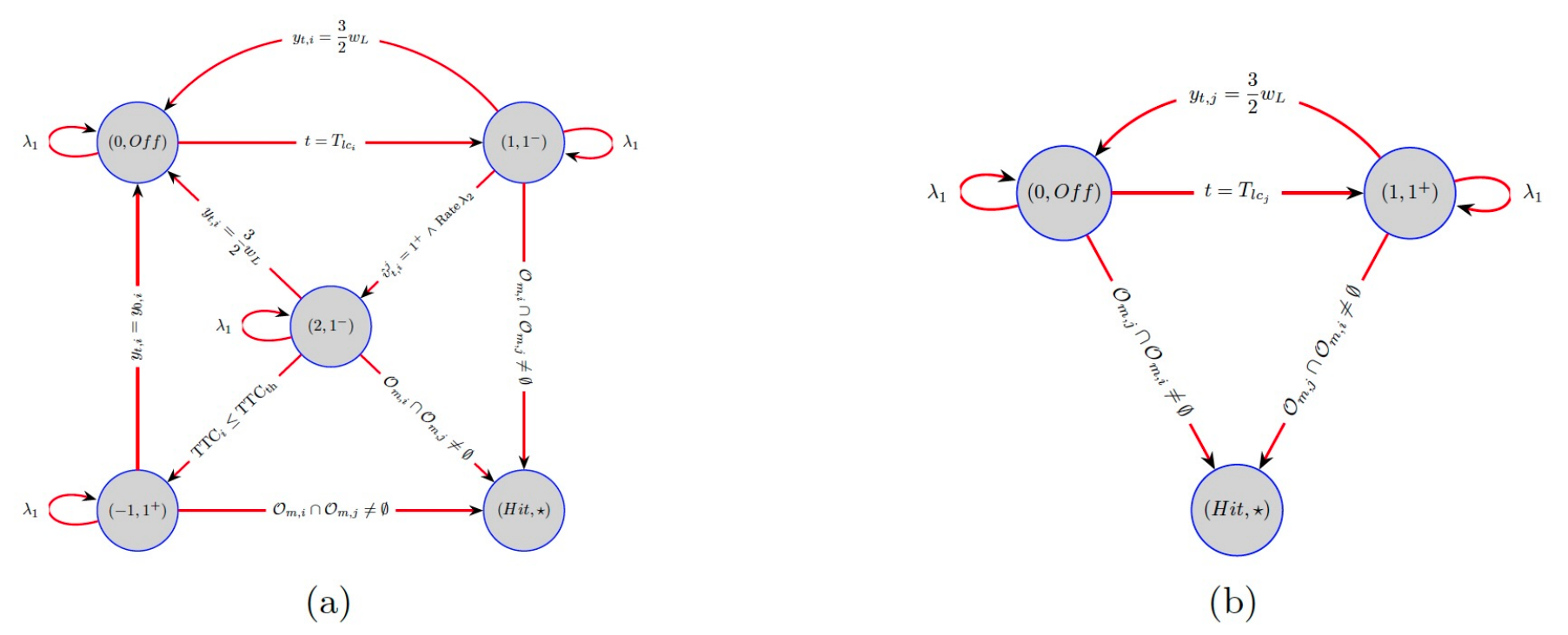}
	\caption{Transition graphs of the GSHS models for AVs $i = ER$ and $j = EL$ are shown in Figures (a) and (b), respectively. In these diagrams, $w_L$ denotes the lane width, while $T_{{lc}_i}$ and $T_{{lc}_j}$ represent the time instants at which AVs $i$ and $j$ initiate their lane-change decisions. The inferred intent $\hat{\upsilon}_{t,i}^j$ is derived from the SA's discrete state.}
\label{fig:Hybrid}
\end{figure}

When the autonomous vehicle $ER$ enters the decision-making mode $(2, 1^-)$, indicating that it is aware of another AV executing a lane change, it evaluates the TTC metric to determine whether to proceed with its own maneuver or abort and return to its original lane—\emph{i.e.}, transition to mode $(-1, 1^+)$. To formalize this decision process, a threshold value $\mathrm{TTC}_{\mathrm{th}}$ is used; if the computed TTC satisfies $\mathrm{TTC}_{ER} \leq \mathrm{TTC}_{\mathrm{th}}$, then completing the lane change is considered unsafe, and $ER$ will revert to its previous lane.
The corresponding transition graphs for the GSHS models of the two AVs, $i = ER$ and $j = EL$,  in this specific scenario are illustrated in Figure~\ref{fig:Hybrid}.

The calculated rare-collision probability using the \emph{interacting particle system-based estimation with fixed assignment splitting (IPS-FAS)} algorithm~\cite{MA2023101303} and the Monte-Carlo (MC) simulation are reported in Table~\ref{tab:results-GSHS} for the sake of comparison. This table underscores the superiority of the IPS-FAS algorithm over MC simulation. While MC yields a zero probability outcome, IPS-FAS provides a probability on the order of $10^{-7}$, highlighting its precision. Given that AVs belong to safety-critical systems, the precision of calculations within their decision-making is of vital importance.
\begin{table*}[h!]
	\centering
	\caption{The value of mean probabilities $\hat \gamma$ for various $\mu_r$ in $\mu_{r_x} = \mu_r \mathcal R_x$ and $\mu_{r_y}  = \mu_r \mathcal R_y$ using IPS-FAS~\cite{MA2023101303} and MC, with $w_L = 3.5\, \text{m}$, $\mu_d = 0.6\,\text{s}$, and $\mathrm{TTC}_{\mathrm{th}} = 10\, \text{s}$.\label{tab:results-GSHS}}
	\begin{tabular}{lccccc}  
		\toprule
		\multirow{2}*{Algorithm} & 
		\multicolumn{5}{c}{$\mu_r$} \\
		\cmidrule(l){2-6}
		{}             & $1.5825$ & $1.6275$ & $1.6725$ & $1.7$ & $1.7375$ \\
		\midrule
		IPS-FAS &  $1.9131 \times 10^{-4}$    & $8.6350\times 10^{-5}$ &  $7.6300 \times 10^{-6}$  & $2.8725 \times 10^{-6}$ & $5.4320 \times 10^{-7}$   \\
		MC      &   $1.8000\times 10^{-4}$   &   $7.9000\times 10^{-5}$    &     $0$     &    $0$    &     $0$     \\
		\bottomrule
	\end{tabular}
\end{table*}

\subsection{Reduced Variants}\label{sub:REDUCED}
To enable a wider range of tools to participate in the benchmark case studies, the system specifications may be simplified to more tractable forms, \emph{e.g.,} reachability, reach-while-avoid, or safety. 
In addition, we introduce the following reduced examples as alternatives, intended for situations in which a tool is unable to handle an actual given benchmark, typically due to challenges such as the state-space explosion problem. We note that tools that discretize the state space often struggle to scale to systems with relatively high dimensionality, highlighting the usefulness of the following reduced alternatives.
\subsubsection{Reduced Patrol Robot}\label{sub:RPR}
Taken from~\cite{wooding2024impactpaper}, the dynamic upon which the robot evolves is
\begin{equation*}
    \begin{bmatrix}
        x_1(k+1)\\
        x_2(k+1)
    \end{bmatrix} = \begin{bmatrix}
        x_1(k) + 10u_1(k)\cos(u_2(k)) + w(k) + \varsigma_1(k)\\
        x_2(k) + 10u_1(k)\sin(u_2(k)) + w(k) + \varsigma_2(k)
    \end{bmatrix}\!\!,
\end{equation*}
where $(x_1,x_2)\in X \coloneqq [-10,10]^2$ represent the spatial coordinate of the location of the robot, $(u_1,u_2)\in U \coloneqq [-1,1]^2$ is the input vector, and $w\in W \coloneqq [-0.5,0.5]$ is the disturbance.
The Gaussian noise $(\varsigma_1,\varsigma_2)$ has the covariance matrix $\Sigma \coloneqq \begin{bmatrix}
    0.75 & 0\\
    0 & 0.75
\end{bmatrix}$. The discrete intervals are respectively $\eta_x = (0.5, 0.5)$, $\eta_u = (0.1, 0.1)$ and $\eta_w = 0.1$. We also define the target set $\mathcal{T} \coloneqq [5,7]^2$ and the avoid set $\mathcal{A} \coloneqq [-2,2]^2$. The guarantees should be found over an infinite horizon.
We consider four scenarios for this case study:
\begin{itemize}
    \item Reach-while-avoid specification with no disturbance;
    \item Reach-while-avoid specification with disturbance;
    \item Reachability specification with no disturbance and $\eta_x = (1, 1)$ and $\eta_u = (0.2, 0.2)$;
    \item Reachability specification with disturbance and $\eta_x = (1, 1)$, $\eta_u = (0.2, 0.2)$.
\end{itemize}
The latter two scenarios are designed to enable fast simulations on standard personal computers.

\subsubsection{Reduced Autonomous Vehicle}\label{sub:AVR}
Borrowed from~\cite{wooding2024impactpaper,reissig2016feedback}, the dynamic that governs a 3-dimensional autonomous vehicle is
\begin{align*}
    \begin{bmatrix} x_1(k+1) \\ x_2(k+1) \\ x_3(k+1)\end{bmatrix}=\begin{bmatrix}
        x_1(k)+\left(u_1(k) \cos \left(\alpha+x_3(k)\right) \cos^{-1} (\alpha)\right) T_s+\varsigma_1(k) \\ x_2(k)+\left(u_1(k) \sin \left(\alpha+x_3(k)\right) \cos^{-1} (\alpha)\right) T_s+\varsigma_2(k) \\ x_3(k)+\left(u_1(k) \tan \left(u_2(k)\right)\right) T_s+\varsigma_3(k)
    \end{bmatrix}\!\!,
\end{align*}
where $\alpha = \arctan (\frac{\tan (u_2)}{2})$, and $T_s = 0.1$ is the sampling time. 
The control inputs $(u_1,u_2)\in U \coloneqq [-1,4]\times[-0.4,0.4]$ represent the wheel velocity and the steering angle, where $\eta_u = (0.5,0.1)$. The state variables $(x_1(k),x_2(k))$ are the spatial coordinates, and $x_3(k)$ is the orientation, where $(x_1,x_2,x_3)\in X \coloneqq [-5,5]^2\times [-3.4, 3.4],$ and $\eta_x = (0.5,0.5,0.2)$.
The Gaussian noise $(\varsigma_1,\varsigma_2,\varsigma_3)$ has the covariance matrix $\Sigma \coloneqq \begin{bmatrix}
    \frac{2}{3} & 0 & 0\\
    0 & \frac{2}{3} & 0 \\
    0 & 0 & \frac{2}{3}
\end{bmatrix}$.
The target and avoid sets $\mathcal{T}$ and $\mathcal{A}$ are, respectively, described by the hyper-rectangles $[-5.75,0.25] \times[-0.25,5.75] \times[-3.45,3.45]$ and $[-5.75,0.25] \times[-0.75,-0.25] \times[-3.45,3.45]$.
For this case study, we consider a reach-while-avoid specification with no disturbances. We additionally consider a reduced complexity version of this case study with $\eta_x = (0.5,0.5,0.4)$ and $\eta_u = (1,0.2)$. The guarantees should be found over an infinite horizon.

\subsubsection{Reduced Building Automation}\label{sub:RBA}
We present a basic $1$-dimensional room temperature system, based on~\cite{ctSSnejati2021compositional}, with both discrete-time and continuous-time dynamics. The \emph{discrete-time} dynamic is provided as
\begin{equation*}
    \Sigma_d\!:x(k+1) = (1 - \beta - \theta(-0.012x(k) + 0.8) )x(k) + \theta T_h(-0.012x(k) + 0.8) + \beta T_e + R\varsigma(k),
\end{equation*}
where $x(k)$ is the temperature of the room, $T_h = 45^{\circ}$C is the heater temperature, $T_e = -15^{\circ}$C is the ambient temperature of the room, $\beta = 0.06$ and $\theta = 0.145$ are conduction factors, and $R=0.1$. The exponential noise $\varsigma$ has a rate of $1$. For this benchmark, we consider a safety specification, in which the state set is $X = [1,50]$, the initial set is $X_{\mathcal I} = [19.5,20]$, and the unsafe sets are $X_{\mathcal U_1} = [1,17]$, $X_{\mathcal U_2} = [23,50].$

On the other hand, the \emph{continuous-time} dynamic is
\begin{equation*}
    \Sigma_c\!:\mathsf{d}{x}(t) = ((-2\eta - \beta - \theta(-0.012x(t) + 0.7))) x(t) + \theta T_h(-0.012x(t) + 0.7) + \beta T_e)\mathsf{d}t + \delta\mathsf{d}\mathbb{W}_t + \rho\mathsf{d}\mathbb{P}_t,
\end{equation*}
where, similar to the discrete-time model, $x$ is the temperature of the room, $T_h = 48^{\circ}$C is the heater temperature, $T_e = -15^{\circ}$C is the outside temperature, and $\eta=0.005$, $\beta = 0.06$, and $\theta = 0.156$ are conduction factors. The system noise consists of both Brownian, with diffusion term $\delta=0.1$, and Poisson process, with reset term $\rho=0.1$ and rate $0.1$. The specification of interest is safety, with the same aforementioned interest regions.

\section{Friendly Competition Results}\label{sec:results}

\subsection{Benchmarking \prtct}
All benchmarks are included in the GitHub repository of \prtct in \texttt{/ex/ARCH-COMP/2025}. For all benchmarks, the minimum confidence level is set as high as possible, and the level set values $\gamma$, $\lambda$, and the value $c$ are then found to achieve the desired confidence. All computations are performed on a Linux machine equipped with an Intel i9-12900 processor, using the \prtct v1.3 release. The summary of results is given in Table~\ref{TAB:Pro}, while additional details for how \prtct was used to solve each benchmark are presented in the sequel.

\begin{table}[h!]
\caption{Summary of the results from running various benchmarks using \prtct. In this table, AS refers to Automated Anesthesia, BA-ct to the continuous-time Building Automation, BA-dt to the discrete-time Building Automation, BA-$4d$ to the four-dimensional Building Automation, VP-dt to the discrete-time Van der Pol, VP-ct to the continuous-time Van der Pol, and IC to the Integrator Chain benchmark.\vspace{0.1cm}}
\label{TAB:Pro}
         \centering
         \begin{tabular}{cccccc}
    \toprule
    Benchmark    & Confidence & Time (seconds) 
            & $\lambda$     & $\gamma$            & $c$            \\
    \midrule
    AS      & $0.95$     & $1.872$      
            & $2.0\times10^{-4}$  
            & $1.01\times10^{-6}$  & $1.25\times10^{-6}$ \\
    BA-ct   & $0.999$    & $0.098$      
            & $3.7\times10^{-2}$  
            & $1.259$             & $2.47\times10^{-6}$ \\
    BA-dt   & $0.999$    & $0.115$      
            & $3.3\times10^{-2}$  
            & $1.128$             & $2.34\times10^{-6}$ \\
    BA-$4d$ & $0.999$    & $2.231$      
            & $4.0\times10^{-3}$  
            & $1.21\times10^{-6}$  & $1.27\times10^{-7}$ \\
    VP-dt   & $0.80$     & $15.29$      
            & $4.0\times10^{-2}$  
            & $6.0\times10^{-3}$   & $4.03\times10^{-4}$ \\
    VP-ct   & $0.80$     & $2.723$      
            & $1.4\times10^{-2}$  
            & $2.0\times10^{-3}$   & $1.23\times10^{-5}$ \\
    IC      & $0.75$     & $32.24$      
            & $7.0\times10^{-3}$  
            & $1.0\times10^{-3}$   & $7.75\times10^{-5}$ \\
    \bottomrule
  \end{tabular}
    \end{table}

\subsubsection{Automated Anesthesia (AS)}

We consider the AS benchmark~\cite[Subsection 2.1]{ARCH18} for a safety \emph{verification} problem, meaning that we assume no control inputs are applied, \emph{i.e.,} $v[k] = 0$ and $\sigma[k] = 0$. This setup effectively assesses the safety of the patient in the absence of external support.
The state space is defined as $X = [0, 7] \times [-1, 11]^2$, with the initial region $X_\mathcal{I} = [4, 6] \times [8, 10]^2$, and the unsafe region $X_\mathcal{U} = X \setminus [1, 6] \times [0, 10]^2$. The specification is evaluated over a time horizon of $10$ seconds. Using \prtct, a degree-$2$ barrier certificate was successfully synthesized.

\subsubsection{Building Automation (BA)}
\prtct can verify safety for both the continuous-time (ct) and discrete-time (dt) reduced versions of the BA benchmark introduced in Subsection~\ref{sub:RBA}. In both cases, barrier certificates of degree~$2$ were successfully synthesized. \prtct can also solve the $4$-dimensional BA benchmark for a safety verification problem (\emph{i.e.,} with $u[k] = 0$) )~\cite[Subsection 4.2]{ARCH19}. The results for this case, using a degree-$2$ barrier certificate, are presented in Table~\ref{TAB:Pro}. The state space is defined as $X = [18, 21]^2 \times [29, 36]^2$, with the initial region $X_\mathcal{I} = [17, 18]^2 \times [8, 10]^2$, and the unsafe regions $X_{\mathcal{U}_1} = [18,18.9]\times [18,21] \times [29, 36]^2$, $X_{\mathcal{U}_2} = [18,21]\times [18,18.9] \times [29, 36]^2$, $X_{\mathcal{U}_3} = [18,21]^2\times [29,29.9] \times [29, 36]$ and $X_{\mathcal{U}_4} = [18,21]^2\times [29,36] \times [29, 29.9]$. We consider a time horizon $T = 10$.

\subsubsection{Van der Pol (VP)}

\prtct can solve both the discrete-time (dt) and the continuous-time (ct) variants of the Van der Pol benchmark for a safety specification. The corresponding results, obtained using degree-$6$ barrier certificates, are presented in Table~\ref{TAB:Pro}. A degree-$6$ barrier certificate was required to achieve the desired confidence level. For both the dt and ct cases we consider; the state space is defined as $X = [-6, 6]^2$, with the initial region $X_\mathcal{I} = [-4.5, 4.5]^2$, and the unsafe regions $X_{\mathcal{U}_1} = [-6,-5]\times [-5,5]$, $X_{\mathcal{U}_2} = [5,6]\times [-5,5]$, $X_{\mathcal{U}_3} = [-5,5]\times [-6,-5]$ and $X_{\mathcal{U}_4} = [-5,5]\times [5,6]$. For dt, we consider a time horizon $T = 5$, while we consider a time horizon $T=13$ for the ct case.

\subsubsection{Integrator Chain (IC)}

We consider the IC benchmark~\cite[Subsection 4.2]{ARCH20} by setting $u[k] = 0$. Results for the benchmark with dimension~$3$ and a barrier certificate of degree~$4$ are presented in Table~\ref{TAB:Pro}. For completeness, this system is described below with $N_s=0.1$:
\begin{align*}
    x_1(k+1) &= x_1(k) + N_sx_2(k)  + \frac{N_s^2}{2}x_3(k) + \frac{N_s^3}{6} + \varsigma_1(k),\\
    x_2(k+1) &= x_2(k) + N_sx_3(k)+ \frac{N_s^2}{2} + \varsigma_2(k),\\
    x_3(k+1) &= x_3(k) + N_s + \varsigma_3(k).
\end{align*}
An independent Gaussian noise distribution describes each variable $\varsigma_i$ with mean $\mu_i = 0$ and variance $\sigma_i=0.1$ for $i\in\{1,2,3\}$. The state space is defined as $X = [-11.0, 11.0]^3$, with the initial region $X_\mathcal{I} = [-10.0, 10.0]^3$, and the unsafe regions $X_\mathcal{U} = X \backslash [-10.1,10.1]^3$. We choose time horizon $T = 5$.

\subsection{Benchmarking \impact}
All benchmarks are included in the GitHub repository of \impact in \texttt{/examples/ARCH-C-} \texttt{OMP/2025}. All computations are performed on a Linux machine equipped with an Intel i9-12900 processor, using the \impact v1.0 release. It is recommended to always use the `sorted' variety of the control synthesis algorithms as they compute much faster. As a general remark on all benchmarks, transitions that lead outside the state space must be treated as transitions to the avoid region to ensure the correctness of IMDP computations. Therefore, an avoid region is specified for reachability tasks, while reach-avoid specifications include additional avoid regions within the state space.

\subsubsection{Automated Anesthesia (AS)}
We consider the AS benchmark~\cite[Subsection 2.1]{ARCH18} as a finite-horizon reachability specification with a time horizon of $10$ seconds. We consider the state space as $[1.0,6.0]\times[0.0,10.0]^2$ and the input space as $[0.0,7.0]\times[0.0,30.0]$. The gridding steps $\eta_x = (0.25, 1, 1)$ and $\eta_u = (1, 30)$ are selected, where the input accounts for both the autonomous control of the machine and the doctor.
Either of the two inputs here could equally has been treated as disturbances, where the system is then robust against the action. The target region is considered $[4.0,6.0]\times[8.0,10.0]^2$.
Accordingly, the total number of states is $2541$, and the number of inputs is $16$. The target set comprises $81$ states as a subset of the state space. The lower and upper bound target vectors each require $25$Mb of memory and take $0.04$ and $1.61$ seconds to compute, respectively. Similarly, the lower and upper bound avoid vectors (probability of leaving the state space) take $0.026$  and $0.035$ seconds to compute. The lower and upper bound transition matrices require $775$Mb of memory and take $0.45$ and $60$ seconds to compute, respectively. Moreover, the controller synthesis using value iteration over $10$ time steps for both the lower and upper bounds takes $3.707$ seconds.

\subsubsection{Building Automation (BA)}
A finite-horizon safety specification is conducted over a time horizon of $6$ seconds, for the $4$d BA system ~\cite[Subsection 4.2]{ARCH19}. We consider the state space as the region $[19.0,20.0]^2\times[30.0,36.0]^2$ and the input space as the region $[17.0,20.0]$. The gridding steps $\eta_x = (0.5, 0.5, 1, 1)$ and $\eta_u = 1$ are considered. 
The total number of states is $1225$, and the number of inputs is $4$. The lower and upper bound avoid vectors (probability of leaving the state space) take $0.007$  and $0.018$ seconds to compute, respectively. The lower and upper bound transition matrices require $48$Mb of memory and take $4.351$ and $4.60$ seconds to compute, respectively. The abstraction is now complete. Controller synthesis using value iteration over $6$ time steps for both the lower and upper bounds takes $0.226$ seconds.

\subsubsection{Integrator Chain (IC)}
A finite-horizon reachability specification is conducted over a horizon of $5$ time steps. We consider a simple $2$-dimensional version of the IC benchmark~\cite[Subsection 4.2]{ARCH20} with $N_s=0.1$:
\begin{align*}
&x_1(k+1)= x_1(k) + N_sx_2(k) + \frac{N_s^2}{2}u_1(k) + \varsigma_1(k),\\
 &x_2(k+1)= x_2(k) + N_su_1(k) + \varsigma_2(k).    
\end{align*}
We consider the state space as $[-10.0,10.0]^2$ and the input space as $[-1.0,1.0]$.
The gridding steps are set as $\eta_x = (0.5, 0.5)$ and $\eta_u = 0.5$. The target region is considered as $[-8.0,8.0]^2$.
Accordingly, the total number of states is $1681$, and the number of inputs is $5$. The target set comprises $1089$ states as a subset of the state space. The lower and upper bound target vectors require $26.0$Mb of memory and take $0.458$ and $0.932$ seconds to compute, respectively. The lower and upper bound avoid vectors (probability of leaving the state space) take similarly $0.002$ and $0.002$ seconds to compute. The lower and upper bound transition matrices require $14.0$Mb of memory and take $0.020$ and $0.504$ seconds to compute, respectively. Controller synthesis using value iteration over $5$ time steps for both the lower and upper bounds takes $0.029$ seconds.

\subsubsection{Autonomous Vehicle (AV)}
An infinite-horizon reach-while-avoid specification is conducted as described in the reduced examples for the AV benchmark in (we omit repeating the regions of interest), with the gridding selected as $\eta_x = (0.5, 0.5, 0.2)$. 
Accordingly, the total number of states is $7938$, and the number of input values is $30$. As subsets of the state space, the target set comprises $2178$ states, and the avoid set contains $198$ states. The lower and upper bound target vectors require $2.9$Gb of memory and take $447$ and $432$ seconds to compute, respectively. Also, the lower and upper bound avoid vectors (including probability of leaving the state space) take $39.0$ and $38.9$ seconds to compute. The lower and upper bound transition matrices require $7.4$Gb of memory and take $853$ and $1151$ seconds to compute, respectively. Controller synthesis using interval iteration~\cite{haddad2018interval}, which guarantees convergence over an infinite horizon within an error bound of $\epsilon = 10^{-6}$, takes $136$ seconds for both lower and upper bounds.

\subsubsection{Patrol Robot (PR)}
\impact can handle the reduced form of the PR benchmark, introduced in Subsection~\ref{sub:RPR} (we omit repeating the regions of interest), for a reachability specification over an infinite horizon.
The total number of states is $1681$, and the number of input values is $441$. The target and avoid sets each comprise $49$ states. The lower and upper bound target vectors require $273$Mb of memory and take $4.4$ and $12.8$ seconds to compute, respectively. Similarly, the lower and upper bound avoid vectors (probability of leaving the state space) take $17.8$ and $12.7$ seconds to compute. The lower and upper bound transition matrices require $8.84$Gb of memory and take $175$ and $570$ seconds to compute, respectively. Controller synthesis using interval iteration~\cite{haddad2018interval}, which guarantees convergence over an infinite horizon within an error bound of $\epsilon = 10^{-6}$, takes $575$ seconds for both lower and upper bounds.

\subsubsection{Package Delivery (PD)}
\impact cannot directly handle the syntactically co-safe linear temporal logic (scLTL) specification of the PD benchmark~\cite[Subsection 4.1]{ARCH22}. However, equivalent behavior can be achieved by designing two separate controllers that are alternated depending on whether a package is currently held. One controller handles delivery from region $p_1$ to $p_3$, while the other governs the return to $p_1$ in the event that the package is dropped. We consider the state space $[-6,6]^2$. An input is not described for this case study, so we arbitrarily define the input set $[-1,1]^2$, with the gridding $\eta_x = (0.5,0.5)$ and $\eta_u = (0.1,0.1)$. The total number of states is then $625$, and the number of input values is $441$.
We describe the synthesis of both controllers below. In both cases, region $p_2$ is treated as an avoid set, as it is assumed to be associated with package loss and should therefore be avoided. The regions are defined as follows: $p_1 =[5,6]\times[-1,1]$, $p_2=[0,1]\times[-5,1]$, and $p_3=[-4,-2]\times[-4,-3]$.

\noindent\textbf{Delivery to $p_3$.}
As subsets of the state space, the target set comprises $39$ states, and the avoid set (region $p_2$) contains $15$ states. The lower and upper bound target vectors require $30.2$Mb of memory and take $0.17$ and $1.26$ seconds to compute, respectively. Similarly, the lower and upper bound avoid vectors (including probability of leaving the state space) take $3.63$ and $3.46$ seconds to compute. The lower and upper bound transition matrices require $1.15$Gb of memory and take $3.8$ and $72.5$ seconds to compute, respectively. Controller synthesis using interval iteration~\cite{haddad2018interval}, which guarantees convergence over an infinite horizon within an error bound of $\epsilon = 10^{-6}$, takes $691$ seconds for both lower and upper bounds. This convergence time is relatively high due to the presence of absorbing states in the synthesis step. Currently, \impact does not perform any prior graph analysis to detect and group such states.

\noindent\textbf{Return to $p_3$.}
The target and avoid sets are again composed of $39$ and $15$ states, respectively. The lower and upper bound target vectors require $30.2$Mb of memory and take $0.15$ and $1.4$ seconds to compute, respectively. The lower and upper bound avoid vectors (including probability of leaving the state space) take $3.86$ and $3.66$ seconds to compute. The lower and upper bound transition matrices require $1.15$Gb of memory and take $4.2$ and $66.6$ seconds to compute, respectively.
Controller synthesis using interval iteration~\cite{haddad2018interval}, with an error bound of $\epsilon = 10^{-6}$, takes $20.3$ seconds for both lower and upper bounds.

\subsection{Benchmarking \intervalmdp}
Problem definitions, including system models and specifications, are provided in the repository \url{https://github.com/Zinoex/ArchCompStochasticModels.jl}. We note that the application of \intervalmdp to the benchmark set can be found in the \texttt{arch-comp/2025} directory of \url{https://github.com/Zinoex/IntervalMDPAbstractions.jl}.
All computations are performed on a machine running Linux Manjaro, equipped with an Intel i7-6700K processor and 16GB of RAM. The experiments are conducted using version~0.4.5 of \intervalmdp and the development version of \intervalmdpabstractions using abstractions to orthogonally decoupled interval Markov decision processes.
When reporting the mean error in the sequel, we concretely mean that, for a set of non-terminal abstract states $\mathcal{Q} = \{q_1, \ldots, q_n\}$ with upper and lower bound probabilities $\hat{V}^\pi(q_i)$ and $\check{V}^\pi(q_i)$ under policy $\pi$, the mean error is defined as:
\[
    \frac{1}{|\mathcal{Q}|} \sum_{i = 1}^n \hat{V}^\pi(q_i) - \check{V}^\pi(q_i).
\]
Since we rely on uniform gridding, all regions are of equal size in the sense of the Lebesgue measure, and thus the mean error maps directly to the concrete system.

\subsubsection{Automated Anaesthesia (AS)}
We consider a fully automated variation of the Anaesthesia system, with a finite-horizon safety specification over $10$ time steps requiring the system to remain within the safe set $[1, 6] \times [0, 10]^2$~\cite{ARCH19}. The region of interest—equal to the safe set—is partitioned into a uniform grid of $(12, 20, 20)$ cells. Additionally, we select three evenly spaced input values from the input space $[0, 7]$, \emph{i.e.,} $\{0, 3.5, 7\}$.
The total number of states is $4800$ ($5733$ including sink states). The abstraction is constructed in $0.034$ seconds and requires $12.75$Mb of memory. Pessimistic value iteration and controller synthesis take $0.77$ seconds, while computation of the upper bound satisfaction probability under the synthesized controller takes $0.18$ seconds.
The maximal non-terminal lower bound is $99.98\%$, and the mean is $34.31\%$. The corresponding mean error is $65.69\%$.

\subsubsection{Building Automation System (BA)}
We have tested both CS1 (four-dimensional) and CS2 (seven-dimensional)~\cite[Subsection 4.2]{ARCH19}. For both benchmarks, the specification requires guaranteeing safety specified as staying within $0.5$ degrees Celsius of a setpoint, $20$ degrees Celsius on the first two dimensions for CS1 and $20$ degrees Celsius for the first dimension on CS2, for $6$ time steps. 

For CS1, we consider the region of interest $[19.5, 20.5]^2 \times [30, 36]^2$, which is partitioned into a uniform grid of $(5, 5, 7, 7)$ cells. We select four evenly spaced input values from the input space $[15, 22]$.
The total number of states is $1225$ ($2304$ including sink states). The abstraction is constructed in $0.023$ seconds and requires $2.37$Mb of memory. Pessimistic value iteration and controller synthesis take $0.11$ seconds, while computation of the upper bound satisfaction probability under the synthesized controller takes $0.03$ seconds.
The maximal non-terminal lower bound is $8.12\%$, and the mean is $6.20\%$. The corresponding mean error is $24.67\%$.

For CS2, there is confusion between implementations about which is the correct system (\emph{e.g.,} whether the dynamics are linear or affine, and whether the dynamics include control). Therefore, for completeness, we include the full set of dynamics here:
\begin{align*}
    x({k+1}) &= Ax(k) + Bu(k) + Q + B_w w(k), \quad w \sim \mathcal{N}(\mu, \Sigma)  ,
\end{align*}
where $\mu = \begin{bmatrix}9&15&500&500&35&35\end{bmatrix}^\top$ and $\Sigma = \mathrm{diag}(1, 1, 100, 100, 5, 5)$.
The matrices are the following (truncated to four places after the decimal point\footnote{Please check \url{https://github.com/Zinoex/ArchCompStochasticModels.jl} for the matrices with full precision.}):
\begin{align*}
    A &= \begin{bmatrix}
       0.9678 & 0    & 0.0036 & 0    & 0.0036 & 0    & 0.0036 \\
       0    & 0.9682 & 0    & 0.0034 & 0    & 0.0034 & 0.0034 \\
       0.0106 & 0    & 0.9494 & 0    & 0    & 0    & 0    \\
       0    & 0.0097 & 0    & 0.9523 & 0    & 0    & 0    \\
       0.0106 & 0    & 0    & 0    & 0.9494 & 0    & 0    \\
       0    & 0.0097 & 0    & 0    & 0    & 0.9523 & 0    \\
       0.0106 & 0.0097 & 0    & 0    & 0    & 0    & 0.9794
    \end{bmatrix}\!\!, \quad
    B = \begin{bmatrix}
        0.0195 \\ 0.0200 \\ 0.0 \\ 0.0 \\ 0.0 \\ 0.0 \\ 0.0
    \end{bmatrix}\!\!,\\
    B_w &= \begin{bmatrix}
        0 & 0 & 0.0000 & 0 & 0.0019 & 0 \\
        0 & 0 & 0 & 0.0000 & 0 & 0.0015 \\
        0.0459 & 0 & 0 & 0 & 0 & 0 \\
        0.0425 & 0 & 0 & 0 & 0 & 0 \\
        0 & 0.0397 & 0 & 0 & 0 & 0 \\
        0 & 0.0377 & 0 & 0 & 0 & 0 \\
        0 & 0 & 0 & 0 & 0 & 0
    \end{bmatrix}\!\!, \quad Q =\begin{bmatrix}
        0.0493 \\ -0.0055 \\ 0.0387 \\ 0.0189 \\ 0.011 \\ 0.0108 \\ 0.0109 
    \end{bmatrix}\!\!.
\end{align*}

We consider the region of interest $[19.5, 20.5] \times [19, 22] \times [18, 22]^5$, which is partitioned into a uniform grid of $(10, 15, 8, 8, 8, 8, 8)$ cells. We select eight evenly spaced input values from the input space $[15, 22]$.
The total number of states is $4\,915\,200$ ($10\,392\,624$ including sink states). The abstraction is constructed in $6.32$ seconds and requires $781.31$Mb of memory. Pessimistic value iteration and controller synthesis take $6396$ seconds, while computation of the upper bound satisfaction probability under the synthesized controller takes $1533$ seconds.
The lower bound for all states is less than $1 \times 10^{-10}$, and the corresponding error exceeds $99.99\%$.

\subsubsection{Van der Pol (VP)}
We have tested the discrete-time variant of the Van der Pol model, with additive control and additive Gaussian noise presented in Problem 5 of \cite{ARCH22}. 
We consider a region of interest equal to the safe set of the specification, $[-5, 5]^2$, which is partitioned into a uniform grid of $(50, 50)$ cells. Ten evenly spaced input values are selected from a subset of the input space $[-1, 1]$. The total number of states is $2500$ ($2601$ including sink states). The abstraction is constructed in $3.69$ seconds and requires $39.35$Mb of memory to store. Pessimistic value iteration and controller synthesis take $16.93$ seconds, while computing the upper bound satisfaction probability with the given controller takes $0.33$ seconds. The maximal non-terminal lower bound is $47.25\%$, and the mean is $29.19\%$. Finally, the mean error is $63.70\%$.

\subsubsection{Patrol Robot (PR)}
We consider the \emph{reduced} Patrol Robot benchmark without disturbances, as introduced in Subsection~\ref{sub:RPR}. 
The region of interest is partitioned into a uniform grid of $(21, 21)$ cells for the reachability problem, resulting in $441$ states ($484$ including sink states), and into $(41, 41)$ cells for the reach-while-avoid problem, resulting in $1681$ states ($1764$ including sink states).
Similarly, input values are selected on a uniform grid of $(11, 11)$ cells for the reachability problem, and $(21, 21)$ cells for the reach-while-avoid problem, over the defined input space.

For the reachability problem, the abstraction is constructed in $0.47$ seconds and requires $36.7$Mb of memory to store. Pessimistic value iteration and controller synthesis take $4.00$ seconds, while computing the upper bound satisfaction probability with the given controller takes $0.002$ seconds. The mean non-terminal lower bound is $99.99\%$. Ultimately, the mean error is $0.007\%$.

For the reach-avoid problem, the abstraction is constructed in $11.02$ seconds and requires $389.1$Mb of memory. Pessimistic value iteration and controller synthesis take $9.16$ seconds, while computing the upper bound satisfaction probability with the given controller takes $0.05$ seconds. The maximal non-terminal lower bound is $99.99\%$, and the mean is $99.98\%$. Also, the mean error is $0.02\%$.

\subsubsection{Automated Vehicle (AV)}
We consider the seven-dimensional variant of the automated vehicle presented in Section~4.3 of~\cite{ARCH20}. 
The region of interest is defined as $[-12, 12]^2 \times [-0.5, 0.5] \times [-2.5, 2.5] \times [-0.35, 0.35] \times [-0.5, 0.5] \times [-0.05, -0.05]$, and is partitioned into a uniform grid of $(6, 6, 5, 5, 7, 5, 5)$ cells. Input values are selected on a uniform grid of $(5, 5)$ cells.
The total number of states is $157\,500$ ($508\,032$ including sink states). The abstraction is constructed in $18\,799$ seconds and requires $3.03$Gb of memory. Pessimistic value iteration and controller synthesis take $89\,761$ seconds, while computation of the upper bound satisfaction probability under the synthesized controller takes $3775$ seconds.
The certified satisfaction probability is trivially zero as no region in the partitioning is fully contained in the target set; \emph{i.e.,} no abstract state can guarantee that the target set is reached (see \cite[Lemma 1]{Cauchi2019HSCC} for more details on the importance of aligning the partitioning and labeling).

\subsection{Benchmarking \syscore}
All computations are performed on a machine with a $2.3$ GHz quad-core Intel Core i5 processor and $16$ GB of $2133$ MHz memory, with reported values averaged over five runs.
\subsubsection{Building Automation System (BAS)}
We have benchmarked BAS for both CS1 (four-dimensional) and CS2 (seven-dimensional)~\cite[Subsection 4.2]{ARCH19}. We have updated the output to be only the temperature in zone one for CS1, so that it is more in line with the seven-dimensional system in CS2. The goal is to synthesize a controller that maintains the temperature in zone 1 at $20 ^{\circ}C$ with a maximum permissible deviation of $\pm 0.5 ^{\circ}C$ over $6$ consecutive time steps.

For CS1, we reduced the model order from four dimensions to two using \texttt{ModelReduction} with setting \texttt{f=0.15}. A finite-state abstraction of the reduced-order model was constructed by gridding the input space with \texttt{lu=3}. We chose to use $50\%$ of the input space for actuation and $30\%$ for feedback. After reducing the state space using \texttt{ReduceX}, we constructed a finite-state abstraction via \texttt{FSabstraction}, yielding a total of \texttt{l=[2000*2000]} grid cells.
To quantify the similarity between the original and reduced models, we used \texttt{QuantifySim} with $\epsilon_1 = 0.1$ and $\epsilon_2 = 0.022$, obtaining $\delta_1 = 0.0054$ and $\delta_2 = 0$. The controlled systems were simulated six times, making sure the output is shifted with respect to the steady-state solution.
Regarding performance improvement due to stochastic model reduction, we observed a peak satisfaction probability of $0.946$ when excluding KK filtering (via \texttt{KKfilter}). Thus, for CS1, our toolset’s accuracy improved by $2.85\%$ with the inclusion of stochastic model reduction, and the $\delta$-error was reduced by $50\%$ (factor 2 improvement). The computation time was not significantly affected.
The breakdown of computation times is as follows: the abstraction step takes $5.6$ seconds, similarity quantification $17.7$ seconds, controller synthesis $16.3$ seconds, and control refinement and deployment $0.40$ seconds, yielding a total computation time of $40.12$ seconds. The total memory requirement is $3174.2$ MB.

For CS2, the number of grid cells in the state space is $9 \times 10^6$. The total computation time is $71.02$ seconds, and the memory usage is $5302.3$ MB. We use the four- and seven-dimensional building automation benchmarks to analyze the scalability of \syscore. We analyze the scalability with respect to the number of states of the abstract model (number of grid cells) and the number of states of the Deterministic Finite Automata (DFA). Note that, since both models are reduced to a two-dimensional state space, they cannot be compared in terms of state space scalability. In Table~\ref{tab:syscore_bas_results}, we list the computation time and memory usage for all experiments. It should be noted that we obtained equivalent accuracy across experiments with the same specification horizon (corresponding to the size of the DFA). When increasing the specification horizon, we observed peak satisfaction probabilities of $0.958$ and $0.847$ for the four-dimensional BAS model (original peak probability of $0.973$), and $0.962$ and $0.954$ for the seven-dimensional BAS model (original peak probability $0.973$).

We can see that the four-dimensional BAS case study scales very well with respect to both the number of grid cells and the size of the DFA, in terms of computation time and memory usage. More specifically, the increase in computation time and memory usage with respect to the DFA size is even smaller.
For the seven-dimensional BAS case study, we conclude that it scales reasonably well with respect to the number of grid cells, although the increase is no longer linear. However, there is a notable outlier in computation time when using a DFA of size 12, indicating that the overall scalability with respect to DFA size requires further investigation and improvement in the next release of \syscore.

\begin{table}[h!]
\caption{The total computation time in seconds (s) and memory usage in megabytes (MB)
for different settings of the building automation system benchmark. Here, dim. and Comp.
are abbreviations for dimension and computation, respectively. The size of the DFA refers
to its total number of states, and Grid lists the total number of states of the abstract model.\vspace{0.1cm}}
\label{tab:syscore_bas_results}
\centering
\begin{tabular}{|l|c|c|c|c|c|}
\hline
 & \textbf{State dim.} & \textbf{DFA size} & \textbf{Grid} & \textbf{Comp. time (s)} & \textbf{Memory (MB)} \\
\hline
\multirow{3}{*}{\textbf{BAS 4D}} & 4 & 8 & $4 \times 10^6$ & 40.12 & 3174.2 \\
 & 4 & 8 & $6 \times 10^6$ & 57.58 & 4763.4 \\
 & 4 & 8 & $8 \times 10^6$ & 75.76 & 6344.61 \\
\hline
\multirow{3}{*}{\textbf{BAS 4D}} & 4 & 8 & $4 \times 10^6$ & 40.12 & 3174.2 \\
 & 4 & 10 & $4 \times 10^6$ & 47.84 & 3430.18 \\
 & 4 & 12 & $4 \times 10^6$ & 59.58 & 3686.16 \\
\hline
\multirow{3}{*}{\textbf{BAS 7D}} & 7 & 8 & $9 \times 10^6$ & 71.02 & 5302.3 \\
 & 7 & 8 & $12 \times 10^6$ & 114.6 & 7071.64 \\
 & 7 & 8 & $16 \times 10^6$ & 258.6 & 9428.11 \\
\hline
\multirow{3}{*}{\textbf{BAS 7D}} & 7 & 8 & $9 \times 10^6$ & 71.02 & 5302.3 \\
 & 7 & 10 & $9 \times 10^6$ & 153.4 & 5878.05 \\
 & 7 & 12 & $9 \times 10^6$ & 776.8 & 6453.83 \\
\hline
\end{tabular}
\end{table}

\subsubsection{Van der Pol (VP)}
We have benchmarked VP to show how \syscore can be applied to nonlinear stochastic systems. We considered the discrete-time dynamics of the Van der Pol oscillator~\cite{ARCH22}, which, for completeness, are given by:
\begin{align}
\begin{array}{l}
x_{1}(k+1)=x_{1}(k)+\tau x_{2} (k)+\varsigma_{1}(k), \\
x_{2}(k+1)=x_{2}(k)+\tau\left(-x_{1}(k)+\left(1-x_{1}(k)^{2}\right) x_{2}(k)\right)+u(k)+\varsigma_{2}(k),
\end{array}
\end{align}
where the sampling time $\tau$ is set to be $0.1$ seconds, $\varsigma \sim \mathcal{N}(0,0.2I_2)$, and $y=x$. We defined the state space ${X}=[-3,3]^2$, the input space ${U}=[-1,1]$, and the output space ${Y}={X}$. 
For the Van der Pol oscillator, the goal is to synthesize a controller such that the system remains in the region $p_1 \coloneqq {X}$ until it reaches the region $p_2 \coloneqq [-1.4, -0.7] \times [-2.9, -2]$, corresponding to the scLTL specification $p_1 \, \cup \, p_2$. We constructed a piecewise-affine (PWA) approximation choosing the parameter $N=[41 \quad 41]$. As the second part of abstraction, a finite-state abstraction (\texttt{sysAbs}) of the PWA approximation was constructed using \texttt{GridInputSpace} and \texttt{FSAbstraction} with \texttt{l=[600 600]} grid cells. To generate a simulation relation between the abstraction and the original model, we set $\epsilon=0.1$ and computed a suitable weighting matrix for the simulation relation. To reduce the computation time, we only used a finite number of states to compute this weighting matrix. The abstraction step takes $1440$ seconds. The similarity quantification takes $1748.6$ seconds. Controller synthesis takes $2.85$ seconds. Control refinement and deployment take $1.42$ seconds. The overall computation time is $3191.6$ seconds, and the memory usage is $178.83$ MB.

\subsubsection{Package Delivery}
We have benchmarked PD to show the capabilities  of \syscore in handling complex scLTL specifications beyond basic reach-while-avoid scenarios, \textit{i.e.,} cyclic DFAs. We consider linear time-invariant (LTI) system
\begin{align}
    \left\{\begin{array}{l}
\hspace{-0.2cm}x(k+1)=A x(k)+B u(k)+ B_{w} \varsigma(k), \\
\hspace{-0.2cm}y(k)=C x(k),
\end{array}\right.
\end{align}
where $A \coloneqq 0.9I_2$, $B \coloneqq I_2$, $B_w \coloneqq \sqrt{0.2}I_2$, $C \coloneqq I_2$, and disturbance $\varsigma\sim\mathcal{N}(0,I_2)$.
We empirically tuned the parameter of state abstraction as $l=[400,400]$ to generate a simulation relation using \texttt{QuantifySim} with an epsilon of $0.075$. We simulated the controlled system using \texttt{ImplementController} for $N = 60$ time steps, starting from the initial state $x_0 = [-5, -5]^\top$. Note that $N$ is an empirical parameter and should be set high enough for the DFA to terminate. The abstraction step takes $1.66$ seconds. The similarity quantification takes $6.19$ seconds. Controller synthesis takes $1.71$ seconds. Control refinement and deployment take $0.70$ seconds. The overall computation time is $11.02$ seconds, and the memory usage is $133.40$ MB. 

\subsection{Benchmarking Guided Simulation in \hpnmg}

\begin{table}[t!]
	\caption{Probabilities for $\varphi_{A_{\text{rare}}}= (m_{P_\text{n}}=0) \ U^{[0,30]}(x_{P_0}\ \geq 0.1) $, estimated error (for \hpnmg) resp.\ confidence interval (for the other tools but \probreach) and run time for the mean duration of raining $\mu=\SI{1.5}{\hour}$ and the capacity of the community buffer $P_\text{c}$ for the water sewage facility benchmark with extension A. \probreach's results are upper bounds of the probability of satisfying $\varphi_{A_{\text{rare}}}$.
	Note that the capacity of the community buffer $P_{\text{c}}$ is given in $10^6$ liters. Simulation parameters: Confidence level $95\%$, aim is to obtain an interval width of $\pm 10\%$ (not possible with HYPEG for $38$, $42$ since run time is too long).
    }
	\label{tab:watersewageArare}
	\centering
	\begin{tabularx}{\linewidth}{|c|Y|Y|Y|Y|Y|Y|}
	\hline
		\multicolumn{1}{|c|}{Par.} &
		\multicolumn{6}{c|}{Tools}                                                                                                  \\
	\hline
	          $P_\text{c}$   &     
	          \hpnmg                                                                                                            & \hypeg                                                                              & \modes MC & \modes Restart & \probreach  &\hpnmg guided sim.\\
	\hline\hline
		           $30$   &
		           \makecell{$1.21\times 10^{-3}$\\\footnotesize$2.3\times 10^{-7}$ \\ \footnotesize $0.142$\,\si{s}} & \makecell{$1.19\times 10^{-3}$\\\footnotesize$\pm 1.0\times 10^{-4}$ \\ \footnotesize $309.5$\,\si{s}} & \makecell{$1.14\times10^{-3}$\\\footnotesize$\pm10\,\%$\\\footnotesize$1$\,\si{s}} &
		           \makecell{$1.15\times10^{-3}$\\\footnotesize$\pm10\,\%$\\\footnotesize$1$\,\si{s}} & \makecell{\small $\leqslant 1.21\mathord{ \times} 10^{-3}$\\\footnotesize$51$\,\si{s}} &\makecell{$1.20\times 10^{-3}$\\\footnotesize$\pm10\,\%$ \\ \footnotesize $1.3$\,\si{s}}\\ 
	\hline
		            $34$    & 
		            \makecell{$1.54\times 10^{-4}$\\\footnotesize$3.1\times 10^{-8}$ \\ \footnotesize $0.133$\,\si{s}}& 
		            \makecell{$1.40\times 10^{-4}$\\\footnotesize$\pm 1.0\times 10^{-5}$ \\ \footnotesize $3314.5$\,\si{s}}
		            & \makecell{$1.49\times10^{-4}$\\\footnotesize$\pm10\,\%$\\\footnotesize$6$\,\si{s}} &
		            \makecell{$1.50\times10^{-4}$\\\footnotesize$\pm10\,\%$\\\footnotesize$6$\,\si{s}} &\makecell{\small $\leqslant 1.48\mathord{\times} 10^{-4}$\\\footnotesize$44$\,\si{s}} &\makecell{$1.58\times 10^{-4}$\\\footnotesize$\pm10\,\%$ \\ \footnotesize $9.9$\,\si{s}}\\
	\hline
		           $38$     &  
		           \makecell{$1.44\times 10^{-5}$\\\footnotesize$2.8\times 10^{-9}$ \\ \footnotesize $0.134$\,\si{s}} &
		           \makecell{$1.55\times 10^{-5}$\\\footnotesize$\pm 5.0\times 10^{-6}$ \\ \footnotesize $1452.8$\,\si{s}}
		           & \makecell{$1.29\times10^{-5}$\\\footnotesize$\pm10\,\%$\\\footnotesize$67$\,\si{s}} &
		           \makecell{$1.34\times10^{-5}$\\\footnotesize$\pm10\,\%$\\\footnotesize$59$\,\si{s}} &\makecell{\small $\leqslant 1.49\mathord{\times} 10^{-5}$\\\footnotesize$37$\,\si{s}} & \makecell{$1.40\times 10^{-5}$\\\footnotesize$\pm10\,\%$ \\ \footnotesize $106.0$\,\si{s}}\\
	\hline
		           $42$     &  
		           \makecell{$9.73\times 10^{-7}$\\\footnotesize$1.1\times 10^{-12}$ \\ \footnotesize $0.106$\,\si{s}} & 
		           \makecell{$9.20\times 10^{-7}$\\\footnotesize$\pm 5.0\times 10^{-7}$ \\ \footnotesize $9426.2$\,\si{s}}
		           & \makecell{$9.45\times10^{-7}$\\\footnotesize$\pm10\,\%$\\\footnotesize$865$\,\si{s}} &
		           \makecell{$1.05\times10^{-6}$\\\footnotesize$\pm10\,\%$\\\footnotesize$705$\,\si{s}} &\makecell{\small $\leqslant 9.15\mathord{\times} 10^{-7}$\\\footnotesize$33$\,\si{s}} & \makecell{$9.31\times 10^{-7}$\\\footnotesize$\pm10\,\%$ \\ \footnotesize $1510.5$\,\si{s}}\\
	\hline
	\end{tabularx}
\end{table}

Within this year's competition, the water sewage facility benchmark has been evaluated additionally by the new method \emph{guided simulation} implemented within the tool \hpnmg. The property $\varphi_{A_{\text{rare}}}$ has been model checked for the water sewage facility with extension~A and compared to results obtained by other tools in 2021~\cite{ARCH21}.
Parameterization and setup are also similar to the report published in 2021. 
The guided simulation was configured similar to the simulation-based tools evaluated in 2021—\hypeg and \modes—to sample a sufficient number of runs to obtain a confidence level of $95\,\%$ with a confidence interval half-width of $10\, \%$ of the estimate.
The guided simulation experiments in \hpnmg were executed single-threaded on a machine equipped with an AMD Ryzen~7 PRO 5850U CPU and $32$GB of RAM.

 \paragraph{Results.}
 Table~\ref{tab:watersewageArare} presents the results and computation times for $\varphi_{A_{\text{rare}}}$ obtained by different tools in 2021, as well as the new results obtained via guided simulation in \hpnmg. 
The confidence interval computed by the guided simulation in \hpnmg overlaps, in all cases, with the confidence intervals produced by the other tools. It can be observed that the guided simulation is significantly faster than \hypeg, which performs \emph{classical} discrete-event simulation on the same hybrid Petri net model. For instance, in the case $P_c = 42$, \hypeg required approximately $9400$ seconds to compute a confidence interval that was twice as large as the one obtained by the guided simulation, which completed in about $1500$ seconds. However, \modes remains faster in both configurations (Monte Carlo and Restart), which we attribute to the model being specifically tuned for the property under consideration. 
In this model, the analytical approach of \hpnmg is also faster than its guided simulation counterpart. This is due to the fact that the model contains only a single random variable. To highlight the advantage of guided simulation, as demonstrated in~\cite{NiehageR25}, greater model complexity and size are required.
Similarly, the recently proposed automated rare-event simulation~\cite{Niehage25NFM} does not bring any benefit in this case study, and hence, further results have not been included.  

\subsection{Benchmarking \amytiss}
Earlier ARCH editions (\textit{e.g.,} \cite{ARCH21}) already provide \amytiss simulation results for a wide range of benchmarks. Consequently, we do not repeat those results here and instead focus on the newly reduced benchmarks described in Subsection~\ref{sub:REDUCED}; the corresponding simulation results are presented in Table~\ref{AMYTISS-RESULTS}. These simulations were conducted on a Linux machine equipped with an Intel Core i7 processor, 16GB of memory, and an Intel Gen12LP HD Graphics NEO GPU.

\begin{table}[t!]
	\centering
	\caption{Benchmark simulation results obtained using \amytiss. Here, $\Phi$ denotes the maximum reachability probability, $\Lambda$ the maximum safety probability, \textsf{RT} the runtime in \emph{seconds}, and N/A indicates values that are not applicable. 
Moreover, superscript~$^{(*)}$ for the Reduced Patrol Robot benchmark refers to the discretization parameters used in the first two items described in Subsection~\ref{sub:RPR}, while superscript~$^{(**)}$ corresponds to the parameters used in the third and fourth items. For all other benchmarks, the discretization parameters are identical to those specified in their respective subsections.}
	\label{AMYTISS-RESULTS}
    \begin{tabular}{@{}llcccc@{}}
		\toprule
		\textsf{Benchmark} & \textsf{Specification} & $\Phi$ &$\Lambda$ & \textsf{RT}\\ 
		\midrule
		\midrule
  	\multirow{2}{*}{Reduced Patrol Robot$^{(*)}$}      & Reach-while-avoid ($w(k)=0$) & $0.81$ & N/A &$2.5$    \\            
& Reach-while-avoid ($w(k)\neq 0$) & $0.64$ & N/A & $14.59$ \\ 
    \hline
 
    \multirow{2}{*}{Reduced Patrol Robot$^{(**)}$}      & Reach-while-avoid ($w(k)=0$) & $0.65$ & N/A &$1.06$    \\            
& Reach-while-avoid ($w(k)\neq 0$) &  $0.47$ &  N/A &$3.01$ \\ 
    \hline
Reduced Autonomous Vehicle     &  Reach-while-avoid & $\approx0.99$ & N/A &$337.1$   \\
		\hline
        	 Reduced Building Automation    &  Safety & N/A & $\approx0.99$ &0.91   \\
             \bottomrule
	\end{tabular}
\end{table}

\section{Conclusion}\label{sec:conclusion}

The 2025 edition of the ARCH-COMP Friendly Competition for Stochastic Models reflects continued progress in formal verification and control synthesis for stochastic systems, with a focus on scalability, usability, and broad applicability. This year's evaluation featured six tools: three newly introduced—\impact, \intervalmdp, and \prtct—and three previously established—\syscore, \hpnmg, and \amytiss. The inclusion of these tools significantly enriched the landscape of methodologies applied across the benchmarks, encompassing a diverse range of approaches such as interval-based abstraction, stochastic barrier certificates, symbolic state-space exploration, and correct-by-construction synthesis.

A key highlight of this edition was the introduction of a new \emph{water distribution network benchmark}, designed to evaluate probabilistic safety guarantees under consumption disturbances. This benchmark represents a practically motivated and structurally rich system, providing an opportunity to test synthesis techniques under stochastic uncertainty and resource constraints. Its inclusion broadens the competition’s coverage of application domains and sets a foundation for further exploration of infrastructure-relevant systems.

To promote broader participation and enhance comparability across tools with varying capabilities, a suite of reduced examples was developed. These simplified variants of existing benchmarks, such as those based on autonomous vehicles, patrol robots, and building automation, are tailored to reduce state-space complexity while preserving core specification challenges (\emph{e.g.,} reachability, safety, and reach-while-avoid). These examples enable tools that struggle with high-dimensional models to still be evaluated meaningfully, thereby supporting a more inclusive and informative comparison.

Additionally, the guided simulation engine newly integrated into \hpnmg was evaluated on the water sewage (WS) benchmark, demonstrating the capability to perform rare-event estimation using stochastic signal temporal logic. This new method leverages symbolic state-space representations and statistical model checking to resolve nondeterminism through reinforcement learning, offering a scalable and automated alternative to traditional model checking for stochastic hybrid systems.

Overall, the 2025 competition has contributed substantial advances in both benchmarking infrastructure and tool capabilities. Moving forward, the stochastic models category will benefit from further refining benchmark specifications, extending tool interoperability, and addressing challenges such as high-dimensional scalability, rare-event estimation, and expressive specification handling. The ongoing collaborative effort within the ARCH community continues to push the boundaries of reliable and automated reasoning for stochastic systems, setting a strong foundation for next year’s competition and future research.

\printbibliography

\end{document}